\documentclass[a4paper]{article}

\usepackage{a4wide}
\usepackage{hyperref}
\usepackage{float}
\usepackage{amsmath}
\usepackage{longtable}

\usepackage{graphicx}
\usepackage{booktabs}
\usepackage{algorithm}
\usepackage{algpseudocodex}
\usepackage[natbib,cref,theorems]{nikis}
\usepackage{url}

\usepackage{anyfontsize}

\usepackage[final]{microtype}

\usepackage{etoolbox}
\newcommand{\zerodisplayskips}{\setlength{\abovedisplayskip}{0pt}\setlength{\belowdisplayskip}{0pt}\setlength{\abovedisplayshortskip}{0pt}\setlength{\belowdisplayshortskip}{0pt}}
          \appto{\normalsize}{\zerodisplayskips}
          \appto{\small}{\zerodisplayskips}
          \appto{\footnotesize}{\zerodisplayskips}

\usepackage{ifthen}
\newboolean{isJournal}
\setboolean{isJournal}{true}
\ifthenelse{\boolean{isJournal}}{\def\Jvar{}
    \def\Jfigs{}
    \newcommand{\J}[2][]{#2}
    \newcommand{\JO}[2][]{#2}
    
    \newcommand{\JFigure}[2][]{#2}
    
    \newenvironment{Jfigure}[1][]{\begin{figure}[#1]}{\end{figure}}
    \newenvironment{Jwrapfigure}[3][]{\begin{wrapfigure}[#1]{#2}{#3}}{\end{wrapfigure}}

}{\usepackage{etoolbox}
    \def\Jvar{}
    \def\Jfigs{}
\newcommand{\J}[2][]{\appto\Jvar{#1#2}}
    \newcommand{\JO}[2][]{#1}
    
    \newcommand{\JFigure}[2][]{\appto\Jfigs{#1#2}}

}

\newcommand*{\DAG}{\instancename{DAG}}
\newcommand*{\Occ}{\ensuremath{\textup{occ}}}
\newcommand*{\OccC}{\ensuremath{\textup{occ}_C}}

\newcommand*{\GCISnep}{GCIS-\instancename{nep}}
\newcommand*{\GCISuni}{GCIS-\instancename{uni}}

\newcommand*{\instancename}[1]{\ensuremath{\mathsf{#1}}} \newcommand*{\functionname}[1]{{\ensuremath{\renewcommand{\rmdefault}{ptm}\fontfamily{ppl}\selectfont\textrm{\textup{#1}}}}} 

\newcommand*{\GST}{\instancename{GST}}

\newcommand*{\Path}{\instancename{path}}

\newcommand*{\ag}{\hat{g}}
\newcommand*{\gLyn}{g_\textup{Lyn}}

\newcommand*{\fnLCE}{\functionname{lce}}
\newcommand*{\fnChild}{\functionname{child}}
\newcommand*{\fnLocate}{\functionname{locate}}
\newcommand*{\fnLookup}{\functionname{lookup}}
\newcommand*{\fnStringDepth}{\functionname{string\_depth}}
\newcommand*{\fnLCA}{\functionname{lca}}

\newcommand*{\typeL}{\texttt{L}}
\newcommand*{\typeS}{\texttt{S}}
\newcommand*{\typeHoshi}{\ensuremath{\texttt{S}^*}}

\newcommand*{\TreeT}{\ensuremath{\mathcal{T}_T}}
\newcommand*{\BunpouT}{\ensuremath{\mathcal{G}_T}}
\newcommand*{\BunpouP}{\ensuremath{\mathcal{G}_P}}
\newcommand*{\Symbols}{\ensuremath{\mathcal{S}}}

\newcommand*{\Takai}[2]{\ensuremath{{#1}^{(#2)}}}

\usepackage{xcolor}
\definecolor{teigiIro}{HTML}{5700B5}
\newcommand*{\teigi}[1]{{\color{teigiIro}\emph{#1}}} 

\begin{document}
\title{Grammar Index By Induced Suffix Sorting
\thanks{This article is an extension of a contribution~\cite{akagi21grammar} to the String Processing and Information Retrieval - 28th International Symposium, SPIRE 2021.
}
}

\author{Tooru~Akagi \and
Dominik~K\"{o}ppl \and
Yuto~Nakashima\and
Shunsuke~Inenaga\and
Hideo~Bannai \and
Masayuki~Takeda
}

\maketitle
\begin{abstract}
  \JO{Pattern matching is the most central task for text indexes.
Many recent indexes leverage compression techniques to 
make pattern matching feasible for massive but highly-compressible datasets.}
\JO[We]{Among this class of indexes, we} propose a new compressed text index built upon a grammar compression
based on induced suffix sorting [Nunes et al., DCC'18].
We show that this grammar exhibits a locality sensitive parsing property, which allows us to specify, given a pattern~$P$, 
certain substrings of~$P$, called \emph{cores}, that are similarly parsed in the text grammar whenever these occurrences are extensible to occurrences of~$P$.
Supported by the cores, 
given a pattern of length~$m$,
we can locate all its \Occ{} occurrences in a text~$T$ of length~$n$ within
\Oh{m \lg |\Symbols| + \OccC \lg|\Symbols| \lg n + \Occ}
time, where 
\Symbols{} is the set of all characters and non-terminals,
$\Occ$ is the number of occurrences, and
$\OccC$ is the number of occurrences of a chosen core~$C$ of~$P$ in the right-hand sides of all production rules of the grammar of~$T$.
Our grammar index requires \Oh{g} words of space and can be built in \Oh{n} time 
using \Oh{g} working space on top of the input text, where $g$ is the sum of the lengths of the right-hand sides of all production rules.
\JO[We practically evaluate that our proposed index excels at locating long patterns in highly-repetitive texts.]{We underline the strength of our grammar index with an exhaustive practical evaluation that gives evidence that our proposed solution excels at locating long patterns in highly-repetitive texts.
}
Our implementation is available at \url{https://github.com/TooruAkagi/GCIS_Index}.
\end{abstract}

\section{Introduction}
\JO{Although highly-repetitive texts are perceived as a common problem instance nowadays due to use cases involving 
large sets of DNA sequences within a very narrow taxonomy group
or huge numbers of slightly different versions of documents maintained in revision control systems,
this problem is difficult to handle due to the large amount of data,
but interesting in the sense that the data is highly compressible.}
Compressed text indexes have become the standard tool for \JO[maintaining highly-repetitive texts]{tackling this problem} when full-text search queries like locating all occurrences of a pattern are of importance.
When working on indexes on highly-repetitive data, a desired property is to have a \emph{self-index}, i.e., a data structure that supports queries on the underlying text without storing the text in its plain form.
Such self-indexes include \emph{grammar indexes}.
A grammar index is an augmentation of the admissible grammar~\cite{kieffer00code} produced by a grammar compressor.
Grammar indexes exhibit strong compression ratios for semi-automatically generated or highly-repetitive texts.
Unlike other indexes that perform pattern matching stepwise character-by-character,
some grammar indexes have locality sensitive parsing properties, which allow them to match certain non-terminals of the admissible grammar built upon the pattern with the non-terminals of the text.
Such a property helps us to perform fewer comparisons, and thus speeds up pattern matching for particularly long patterns,
which could be large gene sequences in a genomic database or source code files in a database maintaining source code.
Here, our focus is set on indexes that support 
$\fnLocate(P)$ queries retrieving the starting positions of all occurrences of a given pattern~$P$ in a given text.

\subsection{Our Contribution}
Our main contribution is the discovery of a locality sensitive parsing property in the grammar produced by the
grammar compression by induced suffix sorting (GCIS)~\cite{nunes18grammar},
which helps us to answer \fnLocate{} with an index built upon GCIS with the following bounds:

\begin{theorem}\label{thmLocatePattern}
Given a text~$T$ of length~$n$, we can compute an indexing data structure on~$T$ in
\Oh{n} time, which can locate all \Occ{} occurrences of a given pattern of length~$m$ in
\Oh{m \lg |\Symbols| + \OccC \lg n \lg |\Symbols| + \Occ} time,
where $\Symbols$ is the set of characters and non-terminals of the GCIS grammar and
$\OccC$ is the number of occurrences in the right-hand sides of the production rules of the GCIS grammar of a selected core of the pattern,
where a core is a string of symbols of the grammar of~$P$ defined in \cref{secCores}.
Our index uses \Oh{g} words of working space,
where $g$ is the sum of the lengths of the right-hand sides of all production rules.
\end{theorem}

Similar properties hold for other grammars such as the signature encoding~\cite{mehlhorn97signature}, ESP~\cite{cormode07esp}, HSP~\cite{fischer20deterministic}, 
the Rsync parse~\cite{gagie19rsync}, or the grammar of \citet[Sect.~4.2]{christiansen21optimaltime}.
A brief review of these and other self-indexes follows.

\begin{table}
\caption{Space (in words) and query time needed for answering \protect\fnLocate{} for a pattern of length~$m$ regarding the indexes addressed in this article, cf.~\cref{secRelatedWork}.}
\label{tabConstruction}
\resizebox{\textwidth}{!}{\newcommand{\gsig}{g_{\text{sig}}}
      \begin{tabular}{l@{\hspace{2em}}*{2}{l}}
	  \toprule
            \multicolumn{1}{c}{Index} &
            \multicolumn{1}{c}{Space} &
            \multicolumn{1}{c}{Locate Time} \\
	    \midrule
	    \citet{claude12grammarindex}
            & ${\Oh{\ag}}$ & ${\Oh{m^2 \lg \lg_{\ag} n + (m + \Occ)\lg\ag}}$ \\
            \citet{gagie12grammar}
            & ${\Oh{\ag + z \lg \lg z}}$ & ${\Oh{m^2 + (m + \Occ) \lg\lg n}}$ \\
            \citet{christiansen21optimaltime}
            & ${\Oh{\gamma\lg(n/\gamma)}}$ & ${\Oh{m + \lg^\epsilon \gamma + \Occ \lg^\epsilon(\gamma\lg(n/\gamma))}}$ \\
            \citet{christiansen21optimaltime}
            & ${\Oh{\gamma\lg(n/\gamma)\lg^\epsilon(\gamma\lg(n/\gamma))}}$ & ${\Oh{m + \Occ}}$ \\
	    Lyndon SLP~\cite{tsuruta20lyndon}
            & ${\Oh{\gLyn}}$ & ${\Oh{m + \lg m \lg n + \Occ \lg \gLyn}}$ \\
            ESP~\cite{takabatake14esp} & ${\Oh{g_{\text{ESP}}}}$ &  ${\Oh{\lg \lg g_{\text{ESP}} (m + \OccC \lg m \lg n) \lg^* n }}$ \\
            Signature~\cite{mehlhorn97signature,nishimoto20dynamic} & ${\Oh{\gsig}}$ & ${\Oh{m \sqrt{\frac{\lg\gsig}{\lg \lg \gsig}}  + \lg \gsig \lg m \lg^* n (\lg n + \lg m \lg^* n) + \Occ \lg n}}$ \\
            GCIS (this work) & ${\Oh{g}}$ & ${\Oh{m \lg |\Symbols| + \OccC \lg n \lg |\Symbols| + \Occ}}$ \\
            $r$-index~\cite{gagie18bwt} & ${\Oh{r \lg(n/r)}}$ & ${\Oh{m + \Occ}}$ \\
	    \bottomrule
        \end{tabular}
      }
	         $n$ is the length of~$T$,
	         $z$ is the number of LZ77~\cite{ziv77lz} phrases of~$T$,
	         $\gamma$ is the size of the smallest string attractor~\cite{kempa18stringattractors} of~$T$,
	 	$\gLyn$ is the size of the Lyndon SLP of~$T$,
	         $\ag$ is the size of a given admissible grammar,
	         $\epsilon > 0$ is a constant,
	         $m$ is the length of a pattern~$P$,
	         $g_{\text{ESP}}$ is the size of the ESP grammar of~$T$,
	         $r$ is number of BWT runs, and
	         $\Occ$ is the number of occurrences of $P$ in $T$.
\end{table}

\subsection{Related Work}\label{secRelatedWork}
\JO{In the field of self-indexes,
grammar compression has won high popularity since a grammar can eliminate repetitions while exhibiting powerful query properties.
A grammar index is a \emph{self-index}, i.e., a data structure that supports queries on the underlying text without storing the text in its plain form.
}

With respect to indexing a grammar for answering \fnLocate{}, 
the first work we are aware of is due to
\citet{claude11grammar} who studied indexes built upon so-called \emph{straight-line programs (SLPs)}.
An SLP is a context-free grammar representing a single string in the Chomsky normal form.
\JO{We present their complexity bounds along with improved versions~\cite{claude12grammarindex,gagie12grammar,claude21grammarcompressed} in \cref{tabConstruction}.}

Other research focused on particular types of grammar,
such as the ESP-index~\cite{maruyama13espindex,takabatake14esp,takabatake16siedm},
an index~\cite{claude16universal} combining Re-Pair~\cite{larsson99repair} with the Lempel--Ziv-77 parsing~\cite{ziv77lz},
a dynamic index~\cite{nishimoto20dynamic} based on signature encoding \cite{mehlhorn97signature},
the Lyndon SLP~\cite{tsuruta20lyndon},
or the grammar index of \citet{christiansen21optimaltime}.
For the experiments in \cref{secExperiments}, we will additionally have a look at other self-indexes capable of \fnLocate{}-queries.
There, we analyze Burrows--Wheeler-transform (BWT)~\cite{burrows94bwt}-based approaches, 
namely the FM-index~\cite{ferragina00fmindex} and the $r$-index~\cite{gagie18bwt}.

Finally, the grammar GCIS has other interesting properties besides being locality sensitive.
\citet{crescenzi03text} used a factorization similar to GCIS to sparsify the text for string matching and indexing in space of the sparsification.
\citet{nunes20grammar} showed how to compute the suffix array and the longest-common-prefix array from GCIS during a decompression step restoring the original text.
Subsequently, \citet{diazdominguez21grammar} showed how to compute the BWT directly from the GCIS grammar,
while \citet{deng22fm} presented a BWT-based index built on the text that has been compressed by the GCIS grammar.
This idea was later used by \citet{hong23acceleration} to accelerate count queries on the FM-index.

\section{Preliminaries}\label{secPreliminaries}

With $\lg$ we denote the logarithm to base two (i.e., $\lg = \log_2$).
Given two integers $i, j$, we denote the interval $[i..j] = \{i,i+1,\ldots,j-1,j\}$, with $[i..j] = \{\}$ if $i > j$.
Our computational model is the standard word RAM with machine word size~$\Om{\lg n}$,
where $n$ denotes the length of a given input string~$T[1..n]$, which we call \teigi{the text},
whose characters are drawn from an integer alphabet~$\Sigma$ of size $n^{\Oh{1}}$. We call the elements of $\Sigma$ \teigi{characters}.
For a string $S \in \Sigma^*$, we denote with $S[i..]$ its $i$-th suffix, and
with $|S|$ its length.
The order $<$ on the alphabet~$\Sigma$ induces a lexicographic order on $\Sigma^*$, which we denote by $\prec$.
A table of all frequently used variable names and their meaning is depicted in \cref{tabSymbols} for reference.

\begin{longtable}{@{}p{.22\linewidth}p{.62\linewidth}p{.12\linewidth}@{}}
\caption{Symbol table listing all used variables and symbols.}
\label{tabSymbols}\\
\toprule
\textbf{Symbol} & \textbf{Meaning} & \textbf{First Appearance} \\
\midrule
\endfirsthead
$n$ & Length of the text $T$ & \cref{secPreliminaries} \\
$T$ & Input text of length $n$ & \cref{secPreliminaries} \\
$\Sigma$ & Integer alphabet & \cref{secPreliminaries} \\
$\Gamma$ & Set of non-terminals & Sect.~\ref{secGCISIndex} \\
$\Symbols$ & Set of characters and non-terminals, $\Symbols := \Sigma \cup \Gamma$ & \cref{secPreliminaries} \\
$m$ & Length of pattern $P$ & \cref{thmLocatePattern} \\
$P$ & Pattern to search for in text $T$ & \cref{thmLocatePattern} \\
$\lg$ & Logarithm to base 2 ($\lg = \log_2$) & \cref{secPreliminaries} \\
$\prec$ & Lexicographic order & \cref{secPreliminaries} \\
$\Occ$ & Number of occurrences of pattern $P$ in text $T$ & \cref{thmLocatePattern} \\
$\OccC$ & Number of occurrences of core $C$ in right-hand sides & \cref{thmLocatePattern} \\
$g$ & Sum of lengths of right-hand sides of all grammar rules & \cref{thmLocatePattern} \\
$\BunpouT$ & GCIS grammar of text $T$, $\BunpouT := (\Sigma, \Gamma, \pi, \Takai{X}{\tau_T})$ & Sect.~\ref{secGCISIndex} \\
$\BunpouP$ & GCIS grammar of pattern $P$ & \cref{secCores} \\
$\pi$ & Production function, $\pi : \Gamma \rightarrow (\Sigma \cup \Gamma)^+$ & Sect.~\ref{secGCISIndex} \\
$\pi^*$ & Expansion function (iterative application of $\pi$) & Sect.~\ref{secGCISIndex} \\
$\TreeT$ & Derivation tree of grammar $\BunpouT$ & Sect.~\ref{secGCISIndex} \\
$\GST$ & Generalized suffix tree & \cref{secGCISIndex} \\
$\DAG$ & Directed acyclic graph representation of grammar & Sect.~\ref{secGCISIndex} \\
$\tau_T$ & Height of derivation tree $\TreeT$ & Sect.~\ref{secGCISIndex} \\
$\tau_P$ & Height of derivation tree of pattern grammar $\BunpouP$ & \cref{secCores} \\
$\Takai{T}{h}$ & String at height $h$ in recursive construction & Sect.~\ref{secGCISIndex} \\
$\Takai{X}{\tau_T}$ & Start symbol of grammar $\BunpouT$ & Sect.~\ref{secGCISIndex} \\
$\Takai{X}{h}_i$ & Non-terminal at height $h$ with index $i$ & Sect.~\ref{secGCISIndex} \\
$\Takai{F}{h}_i$ & Factor $i$ at height $h$ in LMS factorization & Sect.~\ref{secGCISIndex} \\
$\Takai{Y}{h}$ & Non-terminal at height $h$ in matching algorithm & \cref{secExtendMatches} \\
$\Takai{\Gamma}{h}$ & Set of non-terminals at height $h$ & Sect.~\ref{secGCISIndex} \\
$\typeL$ & L-type suffix & Sect.~\ref{secGCISIndex} \\
$\typeS$ & S-type suffix & Sect.~\ref{secGCISIndex} \\
$\typeHoshi$ & LMS suffix (S$^*$-type) & Sect.~\ref{secGCISIndex} \\
$C$ & Core of pattern $P$ & \cref{secCores} \\
$C_p$ & Prefix part of core partitioning & \cref{secCores} \\
$C_s$ & Suffix part of core partitioning & \cref{secCores} \\
$D$ & Array storing extended non-terminals for LCE queries & \cref{secLCEDetail} \\
$W$ & List of non-terminals with pattern occurrences & \cref{secExtendMatches} \\
$R$ & Concatenation of all right-hand sides & \cref{secGCISIndex} \\
$Q$ & Array of starting positions in $R$ & \cref{secPractice} \\
$L$ & Array of expansion lengths & \cref{secGCISIndex} \\
$F$ & Sequence of first symbols (Elias-Fano encoded) & \cref{secPractice} \\
$\Delta$ & Delta-encoded right-hand sides & \cref{secPractice} \\
\fnLCE & Longest common extension function & \cref{secGCISIndex} \\
\fnChild & Child node function in $\GST$ & \cref{secGCISIndex} \\
\fnLocate & Pattern location function & Introduction \\
\fnLookup & Non-terminal lookup function & \cref{secGCISIndex} \\
\fnStringDepth & String depth function for $\GST$ nodes & \cref{secGCISIndex} \\
\fnLCA & Lowest common ancestor function & \cref{secGCISIndex} \\
\bottomrule
\end{longtable}

\subsection{Induced Suffix Sorting}
SAIS~\cite{nong11sais} is a linear-time algorithm for computing the suffix array~\cite{manber93sa}.
We briefly review the parts of SAIS important for constructing the GCIS grammar.
SAIS assigns each suffix of length at least two a type, which is either~$\typeL$ or~$\typeS$:
\begin{itemize}
	\item  $T[i..]$ is an \typeL{} suffix if $T[i..] \succ T[i+1..]$, or
	\item  $T[i..]$ is an \typeS{} suffix otherwise, i.e., $T[i..] \prec T[i+1..]$.
 \end{itemize}
Finally, we stipulate that the suffix $T[n..]$ of length one is always type~$\typeS$.
Since it is not possible that $T[i..] = T[i+1..]$, SAIS assigns each suffix a type.
An \typeS{} suffix $T[i..]$ is additionally an \typeHoshi{} suffix (also called LMS suffix in~\cite{nong11sais})
if $T[i-1..]$ is an \typeL{} suffix.
The substring between two succeeding $\typeHoshi$ suffixes is called an \teigi{LMS substring}. 
In other words, a substring $T[i..j]$ with $i < j$ is an LMS substring if and only if $T[i..]$ and $T[j..]$ are \typeHoshi{} suffixes and 
there is no $k \in [i+1..j-1]$ such that $T[k..]$ is an \typeHoshi{} suffix.
Regarding the defined types, we make no distinction between suffixes and their starting positions
(e.g., the statements that (a) $T[i]$ is type~\typeL{} and (b) $T[i..]$ is an \typeL{} suffix are equivalent).
In fact, we can determine \typeL{} and \typeS{} positions solely based on their succeeding positions with the equivalent definition:
\begin{itemize}
	\item if $T[i] > T[i+1]$, then $T[i]$ is \typeL{};
	\item if $T[i] < T[i+1]$, then $T[i]$ is \typeS{}; 
	\item finally, if $T[i] = T[i+1]$, then $T[i]$ has the same type as $T[i+1]$.
\end{itemize}

In what follows, we define a factorization that splits the text into factors at each $\typeHoshi$ position.
To have the property that each factor starts with an $\typeHoshi$ position, 
we prepend a special character \texttt{\#} smaller than all characters appearing in $T$ to the text $T$.
Then splitting $\texttt{\#} T$ at all $\typeHoshi$ positions gives the factorization $\texttt{\#} T = F_1 \cdots F_z$ such that each factor starts with a \typeHoshi{} position and 
all \typeHoshi{} positions of $T$ are factor starting positions.
This factorization is called \teigi{LMS-factorization}.
The LMS-factorization is uniquely defined by the fact that the \typeHoshi{} positions are the factor beginning positions.
By replacing each factor~$F_i$ by the lexicographic rank of its respective LMS substring\footnote{For SAIS to work, it uses a slightly different order on the LMS substrings, called LMS-order. It differs from the lexicographic order when comparing two LMS substrings, where one of them is a prefix of the other. In such a case, the LMS-order would give the longer string a smaller rank.}, 
we obtain a string $\Takai{T}{1}$ of these ranks.
We recurse on $\Takai{T}{1}$ until we obtain a string $\Takai{T}{\tau_T-1}$ whose rank-characters are all unique or whose LMS-factorization consists of at most two factors.

\subsection{Constructing the Grammar}

We assign each computed factor $\Takai{F}{h}_j$ a non-terminal~$\Takai{X}{h}_j$ such that $\Takai{X}{h}_j \rightarrow \Takai{F}{h}_j$, but omit the delimiter~\texttt{\#}.
The order of the non-terminals $\Takai{X}{h}_j$ is induced by the lexicographic order of their respective LMS-substrings.
We now use the non-terminals instead of the lexicographic ranks in the recursive steps.
If we set $\Takai{X}{\tau_T} \rightarrow \Takai{T}{\tau_T-1}$ as the start symbol, 
we obtain a context-free grammar $\BunpouT := (\Sigma, \Gamma, \pi, \Takai{X}{\tau_T})$,
where $\Gamma$ is the set of non-terminals and 
a function $\pi : \Gamma \rightarrow (\Sigma \cup \Gamma)^+$ that applies (production) rules.
For simplicity, we stipulate that $\pi(c) = c$ for $c \in \Sigma$.
Let $g$ denote the sum of the lengths of the right-hand sides of all grammar rules.
We say that a non-terminal ($\in \Gamma$) or a character ($\in \Sigma$) is a \teigi{symbol}, 
and denote the set of characters and non-terminals with $\Symbols := \Sigma \cup \Gamma$.
We understand $\pi$ also as a string morphism $\pi : \Symbols^* \rightarrow \Symbols^*$ by applying $\pi$ on each symbol of the input string.
This allows us to define the \teigi{expansion}~$\pi^*(X)$ of a symbol~$X$, 
which is the iterative application of $\pi$ until obtaining a string of characters, i.e., $\pi^*(X) \in \Sigma^*$ and $\pi^*(\Takai{X}{\tau_T}) = T$.
Since $\pi(X)$ is deterministically defined, we call $\pi(X)$ the \emph{right-hand side} (\emph{RHS}) of~$X$,
and identify a non-terminal $X$ with its rule $X \to \pi(X)$ when convenient.

\begin{lemma}[\cite{nunes18grammar}]\label{lemGCISconstruct}
  The GCIS grammar \BunpouT{} can be constructed in \Oh{n} time.
  $\BunpouT$ is \emph{reduced}, meaning that we can reach all non-terminals of $\Gamma$ from $\Takai{X}{\tau_T}$.
\end{lemma}

$\BunpouT$ can be visualized by its derivation tree~\TreeT{},
which has $\Takai{X}{\tau_T}$ as its root.
Each rule $\Takai{X}{h}_k \rightarrow \Takai{X}{h-1}_{i} \cdots  \Takai{X}{h-1}_{j}$ defines a node~$\Takai{X}{h}_k$ having $\Takai{X}{h-1}_{i}, \ldots, \Takai{X}{h-1}_{j}$ as its children.
The height of \TreeT{} is $\tau_T = \Oh{\lg n}$ because 
the number of LMS substrings of $\Takai{T}{h}$ is at most half of the length of $\Takai{T}{h}$ for each recursion level~$h$.
The leaves of \TreeT{} are the terminals at height~$0$ that constitute the characters of the text~$T$.
Reading the nodes on height~$h \in [0..\tau_T-1]$ from left to right gives $\Takai{T}{h}$ with $\Takai{T}{0} = T$.
We use \TreeT{} only as a conceptual construct since it would take \Oh{n} words of space.
Instead, we merge (identical) subtrees of the same non-terminal together to form a directed acyclic graph~$\DAG$,
which is implicitly represented by $\pi$ as follows:

By construction, each non-terminal appears exactly in one height of \TreeT{}.
We can therefore separate the non-terminals into the sets $\Takai{\Gamma}{1}, \ldots, \Takai{\Gamma}{\tau_T}$ such that a non-terminal of height~$h$ belongs to $\Takai{\Gamma}{h}$.
More precisely, $\pi$ maps a non-terminal on height~$h \ge 1$ to a string of symbols on height~$h-1$ (in particular: a string of characters for $h=1$).
Hence, the grammar is acyclic.

\J{\JO[\section{Incompressible Example}]{}
 Unfortunately, there are strings with $g = \Ot{n}$, so for a particular input, the grammar does not compress at all:
 Let $T = \prod_{i=0}^m \texttt{a}^i \texttt{b} = \texttt{b} \cdot \texttt{ab} \cdot \texttt{aab} \cdot \texttt{aaab} \cdots$
 with $\Sigma = \{\texttt{a},\texttt{b}\}$ and $\texttt{a} < \texttt{b}$.
 Then we have the rules $\Takai{X}{1}_i \rightarrow \texttt{a}^i \texttt{b}$ for $i \in [0..m]$,
 with $\Takai{X}{2} \rightarrow \prod_{i=0}^m \Takai{X}{1}_i$ being the (production) rule of the start symbol~$\Takai{X}{2}$.
Hence, $\tau_T = 2$ and $g = |T| + |\pi(\Takai{X}{2})| = |T|+ m+1$.
Nevertheless, the experiments in \cref{secExperiments} show that the grammar is suitable for most highly-repetitive text collections.
}

\section{GCIS Index}\label{secGCISIndex}
In what follows, we want to show that we can augment \BunpouT{} with auxiliary data structures for answering \fnLocate{}.
Our idea stems from the classic pattern matching algorithm with the suffix tree, cf.\ the textbook solution of~\citet[APL1]{gusfield97algorithms}. 
The key difference is that we search the core of a pattern --- we defer the definition of a core to \cref{secCores} --- in the RHSs of the rules of \BunpouT{} instead of searching the pattern itself in the text~$T$.
For that, we make use of the generalized suffix tree~\GST{} built upon the RHSs of all rules separated by a special delimiter symbol \texttt{\$} being smaller than all symbols.
Specifically, we rank the rules such that $\{X_1, \ldots, X_{|\Gamma|}\} = \Gamma$ (this ranking will be fixed later), and 
set $R := \pi(X_1) \texttt{\$} \pi(X_2) \texttt{\$} \cdots \pi(X_{|\Gamma|}) \texttt{\$}$.
Since we have a budget of \Oh{g} words, we can afford to use a plain pointer-based tree topology.
The leaf~$\lambda$ with the string label $Y \cdots \texttt{\$} \pi(X_{j+1}) \texttt{\$} \pi(X_{j+2}) \texttt{\$} \cdots \pi(X_{|\Gamma|}) \texttt{\$}$ for $Y$ being a non-empty (but not necessarily proper) suffix of $\pi(X_j)$ stores a pointer to the non-terminal~$X_j$ and an offset~$o$ such that $\pi(X_j)[o..]$ is a prefix of $\lambda$'s string label.
Next, we need the following operations on \GST{}:
\sitemize*{\item First, $\fnLCA(u,v)$ gives the lowest common ancestor~(LCA) of two nodes~$u$ and~$v$.
    We can augment \GST{} with the data structure of~\citet{buchsbaum98lineartime} in linear time and space in the number of nodes of \GST{}.
    This data structure answers \fnLCA{} in constant time.
  \item Next, $\fnChild(u, c)$ gives the child of the node~$u$ connected to~$u$ with an edge having a label starting with $c \in \Gamma \cup \Sigma \cup \{\$\}$.
Our \GST{} implementation answers \fnChild{} in \Oh{\lg |\Symbols|} time.
For that, each node stores the pointers to its children in a binary search tree with the first symbol of each connecting edge as key.
  \item Finally, $\fnStringDepth(v)$ returns the string depth of a node~$v$, i.e., the length of its string label, which is the string read from the edge labels on the path from the root to~$v$.
    We can compute and store the string depth of each node during its construction.
  }The operation \fnChild{} allows us to compute the \teigi{locus} of a string~$S$, i.e., the highest \GST{} node~$u$ whose string label has $S$ as a prefix,
in $\Oh{|S| \lg |\Symbols|}$ time.
For each $\pi(X)$, we augment the locus~$u$ of~$\pi(X)\texttt{\$}$ with a pointer to~$X$ such that we can perform
  $\fnLookup(S)$ for a string of symbols $S \in \Symbols^*$ returning the non-terminal~$X$ with $\pi(X) = S$ or an invalid symbol~$\bot$ if such an $X$ does not exist. The time is dominated by the time for computing the locus of~$S$.
Finally, all leaves in suffix order are stored in a linked list such that we can traverse the leaves in lexicographic order with respect to their corresponding suffixes.

\paragraph{Linkage to the Grammar}
Each non-terminal $X \in \Gamma$ stores an array $X.P$ of $|\pi(X)|$ pointers to the leaves in \GST{} such that the $X.P[i]$ points to the leaf that points back to $X$ and has  offset~$i$ (its string label has $\pi(X)[i..]$ as a prefix).
Additionally, each non-terminal~$X$ stores the length of $\pi(X)$,
an array~$X.L$ of all expansion lengths of all its prefixes, i.e., $X.L[i] := \sum_{j=1}^i |\pi^*(\pi(X)[j])|$,
and an array~$X.R$ of the lengths of the RHSs of all its prefixes, i.e., $X.R[i] := \sum_{j=1}^i |\pi(\pi(X)[j])|$.

\paragraph{LCE queries}
Each internal node~$v$ stores a pointer to the leftmost leaf in the subtree rooted at~$v$.
With that we can use the function
$\fnLCE(X, Y, i, j)$ returning the \teigi{longest common extension} (LCE) of $\pi(X)[i..]$ and $\pi(Y)[j..]$ for $X,Y \in \Gamma$ and $i \in [1..|\pi(X)|], j \in [1..|\pi(Y)|]$.
We can answer $\fnLCE(X, Y, i, j)$ by selecting the leaves~$X.P[i]$ and $Y.P[j]$, retrieve the LCA~$\fnLCA(X.P[i], Y.P[j])$ of both leaves, and take its string depth,
all in constant time.
More strictly speaking, we return $\min(|\pi(X)[i..]|$, $|\pi(Y)[j..]|$, $\fnStringDepth(\fnLCA(X.P[i], Y.P[j])))$, since the delimiter~$\texttt{\$}$ is not a unique character, but appears at each end of each RHS in the underlying string~$R$ of \GST{}.

\paragraph{Complexity Bounds}
\GST{} can be computed in \Oh{g} time~\cite{farach-colton00st}.
The grammar index consists of the GCIS grammar,
\GST{} built upon $|R| = g+|\Gamma|$ symbols, and augmented with a data structure for \fnLCA{}~\cite{buchsbaum98lineartime}.
This all takes \Oh{g} space.
Each non-terminal is augmented with an array $X.P$ of pointers to leaves, $X.L$ and $X.R$ storing the expansion lengths of all prefixes of $\pi(X)$,
which take again \Oh{g} space when summing over all non-terminals.

\section{Pattern Matching Algorithm}
We follow the steps of \citet[Sect.~2]{sahinalp96lca}, 
who identify a substring~$C$ of the parse of the pattern~$P$ that can be matched against the RHSs of the grammar.
They called such a substring~$C$ a \emph{core} of the pattern.
The core needs to exhibit certain properties such that all occurrences of $P$ in $T$ are parsed by the grammar of $T$ in a way that their contained parse of $C$ is identical.
This allows us to retrieve the occurrences of $P$ by 
(1) first finding the occurrences of $C$ in the RHSs of the grammar and then (2) extending these occurrences to occurrences of $P$.
\JO{However, we do not try to find the occurrences of $P$ directly, 
but rather the lowest \DAG{} nodes (a) having~$C$ as a descendant and (b) having such a large expansion that we could extend~$C$ to~$P$.
}

\subsection{Cores}\label{secCores}
Central to our pattern matching algorithm is the concept of \emph{cores}.
A core~$C \in \left(\Takai{\Gamma}{h}\right)^+$ with $\Takai{\Gamma}{0} := \Sigma$ is a string of symbols at a certain height~$h \ge 0$ of the GCIS grammar \BunpouP{} built on the pattern~$P$ with the following properties.
\begin{itemize}
  \item The expansion $\pi^*(C)$ of $C$ is a substring of $P$.
  \item Given an occurrence of $\pi^*(C)$ is covered by consecutive nodes on a height $h \ge 0$ in $\TreeT$ such that every node's expansion covers at least one character of $\pi^*(C)$,
    then this occurrence of $\pi^*(C)$ is \emph{not} part of an occurrence of $P$ in $T$ if these nodes do not have all the same parent node on height~$h+1$.
    Technically speaking, we consider symbols $X_i^{(h)}, \ldots, X_{i+\ell}^{(h)}$ on height~$h$ such that $\pi^*(C)$ is a substring of $\pi^*(X_i^{(h)}\cdots X_{i+\ell}^{(h)})$
    such that $\pi^*(C)$ is a suffix of $\pi^*(X_i^{(h)})$ and a prefix of $\pi^*(X_{i+\ell}^{(h)})$. 
    Given there are non-terminals $X_{j}^{(h+1)}$ and $X_{j'}^{(h+1)}$ being the parents of $X_i^{(h)}$ and $X_{i+\ell}^{(h)}$, respectively, with $j \ne j'$,
    then we cannot extend $\pi^*(X_i^{(h)}\cdots X_{i+\ell}^{(h)})$ to an occurrence of $P$ in $T$.
\end{itemize}
A consequence of the latter property is that for each occurrence~$O$ of~$C$ in~$\TreeT$ whose expansion is contained in an occurrence of~$P$,
this occurrence~$O$ is a (not necessarily proper) substring of the RHS of a rule of \BunpouT{}.

Our definition of \emph{core} is neither constructive nor unique.
In fact, there can be many cores for a given pattern.
We qualify a core by the difference in the number of occurrences of $P$ and $C$ in \TreeT{}.
On the one hand, although a character~$P[i]$ always qualifies as a core, the appearance of $P[i]$ in $T$ is unlikely to be evidence for an occurrence of $P$.
\JO{(Additionally, our algorithm would perform worse than standard pattern matching with the suffix tree when resorting to single characters as cores).}
On the other hand, the non-terminal covering most of the characters of~$P$ might not be a core.
Hence, we aim for the highest possible non-terminal, for which we are sure that it exhibits the core property.

\JFigure{\begin{figure}
\begin{minipage}{0.4\linewidth}
\includegraphics[width=\linewidth]{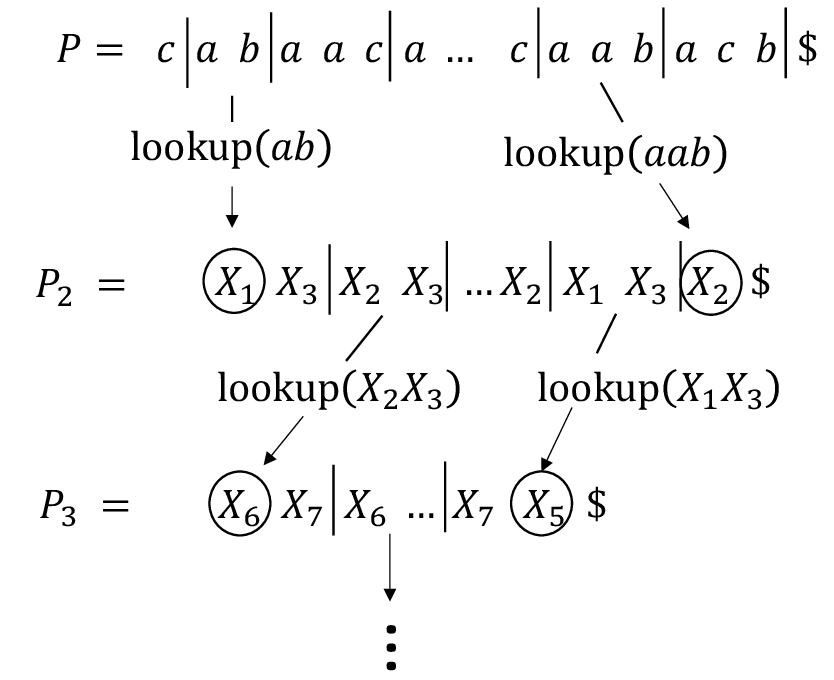}
\end{minipage}
\hfill
\begin{minipage}{0.5\linewidth}
      \caption{Computing \BunpouP{} in \cref{secCores}.
      For each height~$h$, we compute the LMS factorization 
      $\Takai{P}{h} = \Takai{F}{h}_1 \cdots \Takai{F}{h}_z$ of $\Takai{P}{h}$,
      query $\Takai{Y}{h+1}_i = \protect\fnLookup(\Takai{F}{h}_i)$ for $i \in [2..z-1]$,
      and continue with $\Takai{P}{h+1} := \prod_{i=2}^{z_h-1} \Takai{Y}{h+1}_i$
      unless there is one $i$ with $\Takai{Y}{h+1}_i = \bot$, for which we abort the pattern matching.
      }
      \label{figComputePatternGramar}
\end{minipage}
\end{figure}
}

In what follows, we present a greedy heuristic that finds a core~$C$ of~$P$ during the construction of the GCIS grammar~\BunpouP{} of~$P$.
The idea is to discard sufficiently long prefixes and suffixes on each recursion step of the GCIS construction of~$P$, 
allowing us to guarantee the core properties for the remaining center part. 
Since we shave off bordering symbols at each height, the remaining part has the shape of a pyramid, 
where the topmost (and inmost) non-terminal is our core~$C$.

\paragraph{Finding a Core}
We determine a core~$C$ of $P$ during the computation of the GCIS grammar~$\BunpouP$ of~$P$.
During this computation, existing text non-terminals are reused when lookup succeeds, i.e., if $\fnLookup(F) = X$ for a factor~$F$, we reuse the non-terminal~$X$ from \BunpouT{} instead of creating a new non-terminal for~$F$ in \BunpouP{}.
By doing so, we ensure that non-terminals of $\BunpouP$ and $\BunpouT$ are identical whenever their RHSs are equal.
In detail, suppose that we are on height~$h$ and parse~$\Takai{P}{h}$ into LMS factors $\Takai{F}{h}_1 \cdots \Takai{F}{h}_{z_h} = \Takai{P}{h}$.
Next, we retrieve $\Takai{Y}{h+1}_i := \fnLookup(\Takai{F}{h}_i)$ for each $i \in [2..{z_h}-1]$, effectively omitting the first and last factor of the factorization of~$\Takai{P}{h}$.
If one of the \fnLookup{}-queries returns $\bot$, we abort since we can be sure that the pattern does not occur in $T$.
That is because all non-terminals $\Takai{Y}{h+1}_2, \ldots \Takai{Y}{h+1}_{{z_h}-1}$ are cores.
To see this, we observe that prepending or appending symbols to \Takai{P}{h} does not change the factors $\Takai{F}{h}_2, \ldots, \Takai{F}{h}_{z-1} =: \Takai{C}{h}$.
\JO{We give a visual interpretation with \cref{figComputePatternGramar}.}

\paragraph{Correctness}
We show that prepending or appending characters to $\Takai{F}{h}_1$ $\Takai{C}{h}$ $\Takai{F}{h}_{z_h}$ does not modify the computed factorization of $\Takai{C}{h} = \Takai{F}{h}_2, \ldots, \Takai{F}{h}_{z-1}$.
For that, 
we observe that we cannot change the type of any position~$\Takai{C}{h}[i]$ to \typeHoshi{} by prepending or appending characters.
\begin{description}
   \item[Prepending] 
    Firstly, 
    the type of a position (\typeS{} or \typeL{}) depends only on its succeeding position, and hence prepending cannot change the type of a position in $\Takai{C}{h}$.
   \item[Appending] 
    Secondly,
    appending characters can either prolong $\Takai{F}{h}_{z_h}$ or 
    create a new factor~$\Takai{F}{h}_{z_{h+1}}$ since $\Takai{F}{h}_{z_h}$ starts with \typeHoshi{}, and therefore appending cannot change $\Takai{C}{h}$.
\end{description}
An additional insight is that on the one hand, prepending a symbol can only introduce a new factor or extend $\Takai{F}{h}_1$.
On the other hand, appending characters can introduce at most one new \typeHoshi{} position in $\Takai{F}{h}_{z_h}$ that can make it split into two factors. 
We will need this observation later for extending the core to the pattern\JO{, cf.~\cref{figExtendNonTerminals}}.

The construction of $\BunpouP$ iterates the LMS factorization until we are left with a string of symbols $\Takai{P}{\tau_P}$ whose LMS factorization consists of
at most two factors.
In that case, we partition $\Takai{P}{\tau_P}$ into three substrings $C_p \cdot C \cdot C_s$
with $C_p$ and $C_s$ possibly empty, and defined by the two independent rules for $C_p$ and $C_s$. 
\begin{enumerate}
  \item[For~$C_p$:]  If the LMS factorization consists of two non-empty factors $F_1 \cdot F_2$, then $C_p$ is $F_1$.
  \item[For~$C_s$:]    Let $\Takai{P}{\tau_P} = (\Takai{P}{\tau_P}[j_1])^{c_1} \cdots (\Takai{P}{\tau_P}[j_k])^{c_k}$ 
    be the run-length-encoded representation of~$\Takai{P}{\tau_P}$ with $1 = j_1 < \ldots < j_k = |\Takai{P}{\tau_P}|$,
    $c_{i} \ge 1$ for $i \in [1..k]$, and
    $\Takai{P}{\tau_P}[j_i] \not= \Takai{P}{\tau_P}[j_{i+1}]$ for $i \in [1..k-1]$.
    We set  $C_s \gets (\Takai{P}{\tau_P}[j_k])^{c_k}$ if  $\Takai{P}{\tau_P}[j_k] < \Takai{P}{\tau_P}[j_{k-1}]$.
\end{enumerate}
In the other cases, $C_p$ and/or $C_s$ are empty.
    \JO{\Cref{figDetermineCore} visualizes $C_p$ and $C_s$ based on the established rules.}
    While we cannot guarantee that $C_p$ and $C_s$ are cores of~$P$, we can show the following.
    \begin{lemma}
      $C$ is a core of~$P$.
    \end{lemma}
    \begin{proof}
It is left to check the case when $C_s$ is empty. 
The other cases have already been covered by the aforementioned analysis of the cores on the lower heights.
Let $y$ denote the last symbol of $\Takai{P}{\tau_P}$.
If $C_s$ is empty, then $C$ is a suffix of $P$, and as a border case, the last position of $C$ is \typeHoshi{}.
In that case, appending symbols cannot create an LMS boundary inside $C$:
It can change the type of the last position of $C$, 
but the affected boundary is after the last symbol of $C$.
\begin{itemize}
  \item If we append a string $y^\ell x$ with $x < y$ and $\ell \ge 0$ to $C_p C$, 
    then the type of the last position of~$C$ changes to \typeL{}. 
  \item If we append a string $y^\ell z$ with $z > y$ and $\ell \ge 0$, then the last position of~$C$ becomes \typeS{}, but does not become \typeHoshi{} 
    since $\Takai{P}{\tau_P}[j_k] > \Takai{P}{\tau_P}[j_{k-1}]$ due to construction (otherwise $C_s$ would not be empty).
\end{itemize}
    \end{proof}
In particular, if $C_p$ or $C_s$ is empty, then the expansion of the core $C$ is a prefix or a suffix of the pattern, respectively.

\JFigure{\begin{figure}
\begin{minipage}{0.5\textwidth}
\includegraphics[width=\linewidth]{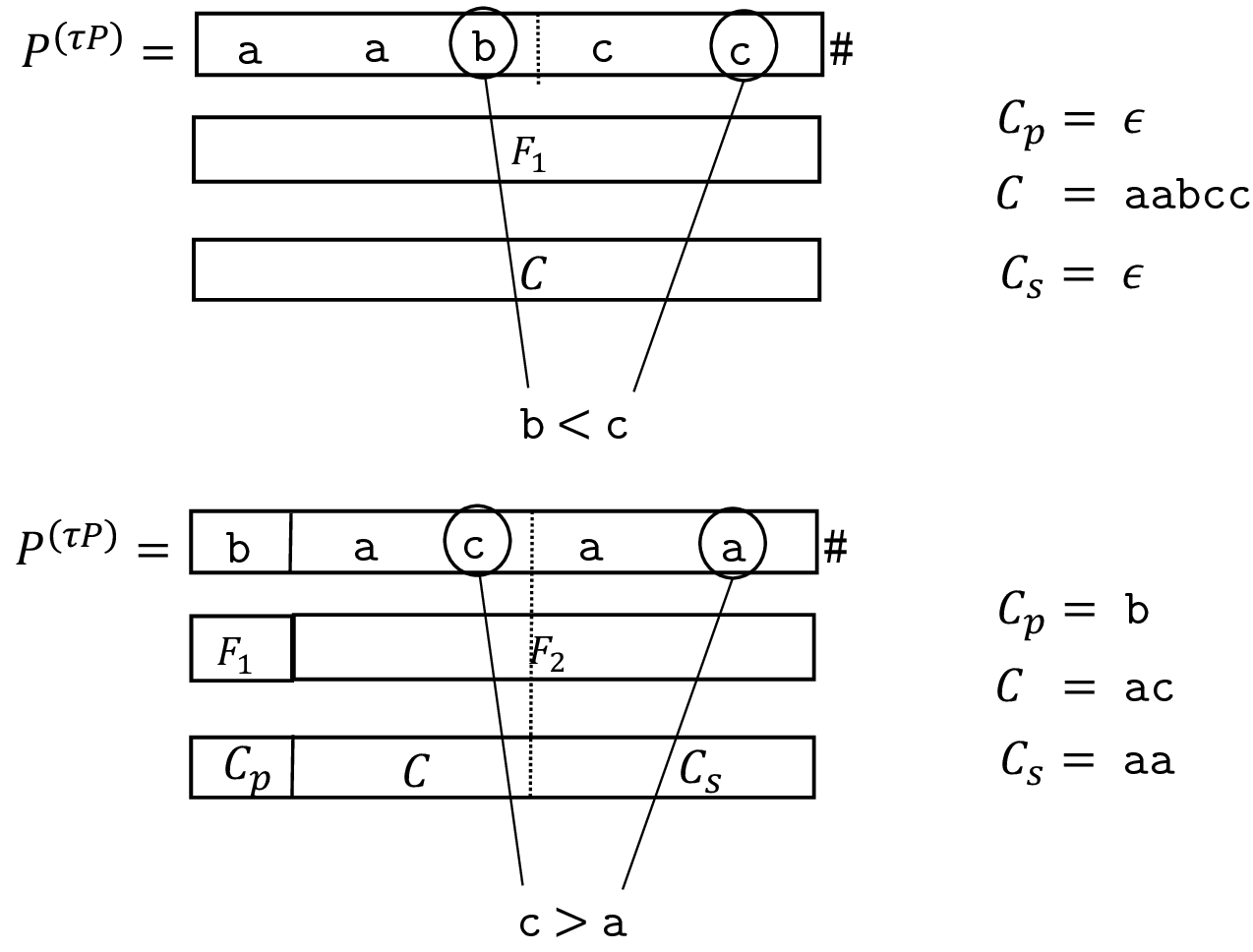}
\end{minipage}
\hfill
\begin{minipage}{0.45\textwidth}
      \caption{Determining the core~$C$ of~$P$ from $\Takai{P}{\tau_P}$.
      Top: The LMS factorization of $\Takai{P}{\tau_P}$ consists only of one factor~$F_1$, so $C_p$ is empty. 
      The last character run~\texttt{cc} consists of a character larger than the previous run, which consists of \texttt{b}'s, such that $C_s$ is empty, too.
      Bottom: The factorization consists of two factors~$F_1 F_2$, so $C_p = F_1$. Since the last character run~\texttt{aa} is lexicographically smaller than the previous one with~\texttt{c}, $C_s$ is set to this last character run.
      }
      \label{figDetermineCore}
\end{minipage}
\end{figure}
}

We can also extract more cores, but for these we can only guarantee
that they have a lower height than $C$.
To put in technical terms, there are symbols $\Takai{A}{1}, \ldots, \Takai{A}{\tau_P-2}$ and $\Takai{S}{1}, \ldots, \Takai{S}{\tau_P-2}$
such that 
\begin{equation}\label{eqCores}
  \begin{aligned}
    D_p :&= \Takai{F}{1}_1 \cdots \Takai{F}{\tau_P-2}_1 \Takai{A}{1} \cdots \Takai{A}{\tau_P-2} \\
    D_s :&= \Takai{S}{\tau_P-2} \cdots \Takai{S}{1} \Takai{F}{\tau_P-2}_{z_{\tau_P-2}} \cdots \Takai{F}{1}_{z_1}\\
    P &= \pi^*( D_p C D_s ) \text{~with~} \pi^*(D_p) = \pi^*(C_p) \text{~and~} \pi^*(D_s) = \pi^*(C_s),
  \end{aligned}
\end{equation}
and $\Takai{A}{h}$, $\Takai{S}{h} \in \Takai{\Gamma}{h}$ are cores of~$P$,
while $\Takai{F}{h}_1$, $\Takai{F}{h}_{z_h} \in (\Takai{\Gamma}{h-1})^*$ 
are factors, for each height~$h \in [1..\tau_P-2]$.

\subsection{Matching with \GST{}}\label{secExtendMatches}
\begin{figure}[t]
\begin{minipage}{0.4\textwidth}
\includegraphics[width=\linewidth]{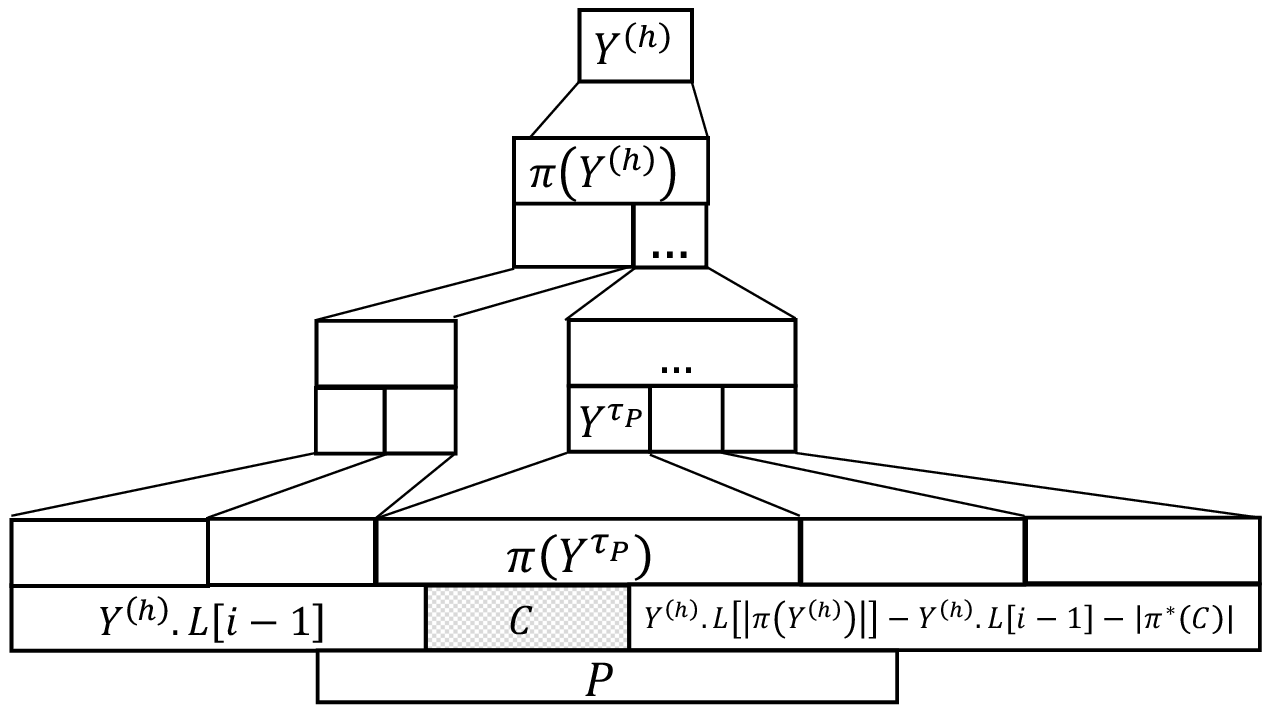}
\end{minipage}
\hfill
\begin{minipage}{0.57\textwidth}
      \caption{Deriving a non-terminal~$\Takai{Y}{h}$ with $\pi^*(\Takai{Y}{h})$ containing~$P$ from
      a non-terminal~$\Takai{Y}{\tau_P}$ with $\pi^*(\Takai{Y}{\tau_P})$ containing $\pi^*(C)$.
      The expansion of none of the descendants of $\Takai{Y}{h}$ towards $\Takai{Y}{\tau_P}$ is large enough for extending its contained occurrence of $\pi^*(C)$ to an occurrence of~$P$. We can check the expansion lengths of the substrings in $\pi(\Takai{Y}{h})$ via the array $\Takai{Y}{h}.L$.
}
      \label{figFindingCoreNonTerminal}
\end{minipage}
\end{figure}

Having $C$, we now switch to \GST{} and use it to find all \DAG{} parents of~$C$,
whose number we denote by $\OccC \in \Oh{g}$. 
This number is also the number of occurrences of~$C$ in the RHSs of all rules of~$\BunpouT$.
Having these parents, we want to find all lowest \DAG{} ancestors of~$C$ whose expansions are large enough to not only cover~$C$ but also $P$ by extending $C$ to its left and right side ---
see \cref{figFindingCoreNonTerminal} for a sketch.
We proceed as follows: We first compute the locus~$v$ of~$C$ in \GST{} in \Oh{|C| \lg |\Symbols|} time via \fnChild{}.
Subsequently, we take the pointer to the leftmost leaf in the subtree rooted at~$v$, and then process all leaves in this subtree by using the linked list of leaves.
For each such leaf~$\lambda$, we compute a path in form of a list~$\lambda_L$ from the non-terminal containing~$C$ on its RHS up to an ancestor of it that has an expansion large enough to cover~$P$ if we would expand the contained occurrence of~$C$ to~$P$.
We do so as follows:
Each of these leaves stores a pointer to a non-terminal~$X$ and a starting position~$i$ such that we know that $\pi^*(X)[i..]$ starts with $\pi^*(C)$.
By knowing the expansion lengths $X.L[|\pi(X)|]$, $X.L[i-1]$, and $|\pi^*(C)|$, we can judge whether the expansion of $X$ has enough characters to be able to extend its occurrence of $C$ to~$P$. 
If it has enough characters, we put $(X, i)$ onto $\lambda_L$ such that we know that $\pi^*(X)[X.L[i-1] + 1..]$ has $C$ as a prefix.
If $X$ does not have enough characters,
we exchange $C$ with $X$ and recurse on finding a non-terminal with a larger expansion.
By doing so, we visit at most $\tau_T = \Oh{\lg n}$ non-terminals per occurrence of~$C$ in the RHSs of $\BunpouT$.
We perform all operations in $\Oh{\OccC \tau_T \lg |\Symbols|}$ time because we query \fnChild{} in every recursion step.

\paragraph{Grammar Subtrees Containing $P$}
The previous step computes, for each accessed leaf~$\lambda$, a list~$\lambda_L$ containing a \DAG{} path $(\Takai{Y}{h^*},\ldots,\Takai{Y}{\tau_P})$ of length~$\Oh{\tau_T}$ and an offset~$\Takai{o}{\tau_P}$ such that 
\begin{itemize}
  \item $|P| \le |\pi^*(\Takai{Y}{h^*})|$, 
  \item $\Takai{Y}{h^*}$ is an ancestor of $\Takai{Y}{\tau_P}$, and
  \item $\Takai{Y}{\tau_P}[\Takai{o}{\tau_P}..]$ starts with $C$.
\end{itemize}
By construction, these paths cover all occurrences of $C$ in \TreeT{}.
We process the \DAG{} node $\Takai{Y}{\tau_P}$ (with different offsets~$\Takai{o}{\tau_P}$) as many times as~$C$ occurs in $\pi(\Takai{Y}{\tau_P})$.
In what follows, we try to expand the occurrence of $C$ captured by $\Takai{Y}{\tau_P}$  and $\Takai{o}{\tau_P}$ to an occurrence of~$P$.
\JO{See \cref{figExtendNonTerminals} for a sketch of the setting.}
Naively, we would walk down from $\Takai{Y}{\tau_P}[\Takai{o}{\tau_P}]$ to the character level and extend the substring $\pi^*(C)$ in both directions by character-wise comparison with $P$.
However, this would take $\Oh{\OccC m \tau_T}$ time since such a non-terminal~$\Takai{Y}{h^*}$ is of height~$\Oh{\tau_T}$.
Our claim is that we can perform the computation in  \Oh{m + \OccC \tau_T} time with the aid of \fnLCE{} and an amortization argument.

\begin{figure}
\includegraphics[width=0.4\linewidth]{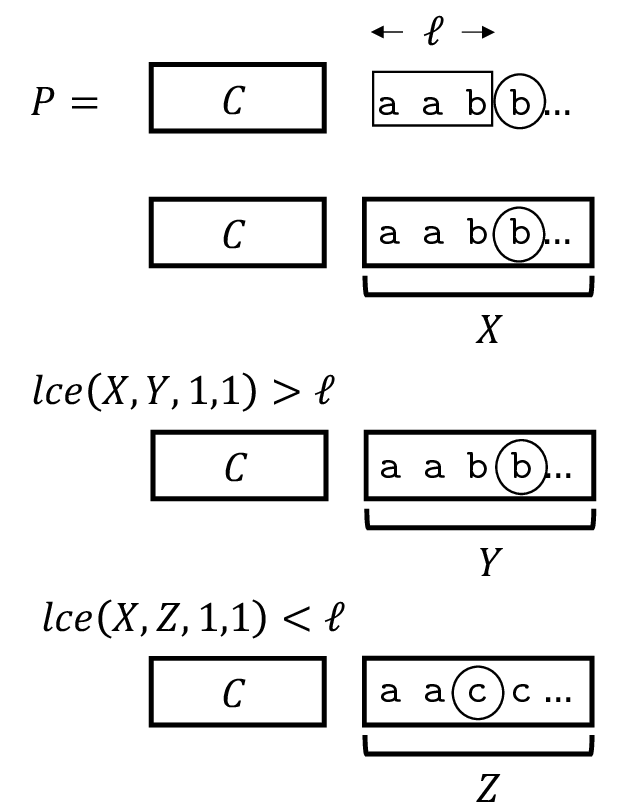}
\hfill
\includegraphics[width=0.4\linewidth]{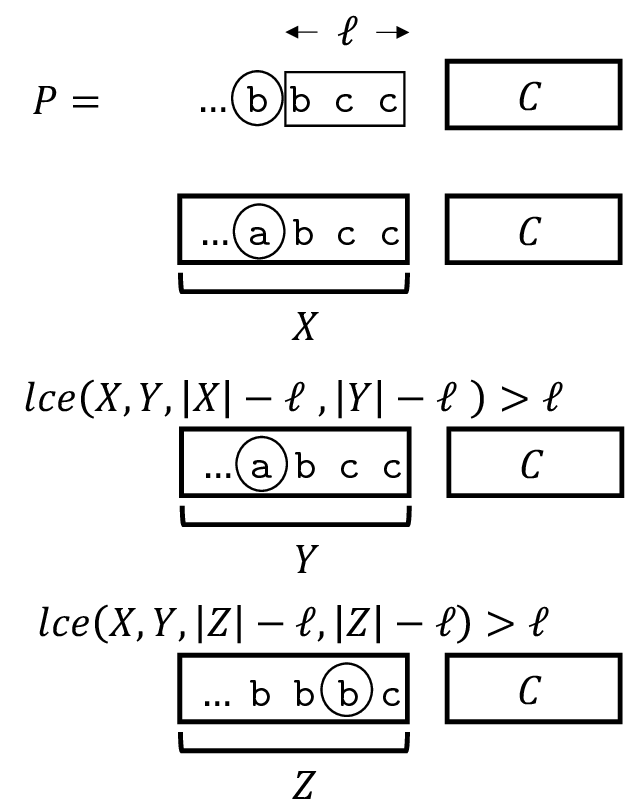}
      \caption{LCE queries issued in \cref{secExtendMatches}.
      Left: Let $X,Y,Z \in \Takai{\Gamma}{h}$.
      Assume that the expansion of the first $\ell$ symbols of the non-terminal~$X$ coincides with the expansion of $C_s$, and that~$X$ is currently stored in $D$.
      To compute the number of matching characters of the expansion of $C_s$ with the expansion of $Y$ or $Z$, we first issue LCE queries with $X$ to decide whether there is a mismatch at one of the first $\ell$ symbols; in such a case we know that we cannot find an occurrence of the pattern, as it is the case with $Z$.
      Right: Symmetric application for extending $C$ to the left.
      The LCE queries compare symbols in RHSs of the non-terminals.
}
      \label{figComputeLCE}
\end{figure}

For that, we use \cref{eqCores},
which allows us to use LCE queries in the sense that we can try to extend an occurrence of $C$ with an already extended occurrence (that may not match $P$ completely) from a different leaf~$\lambda$ and hence a possibly different path (such an extension is safe since we retrieve a substring of consecutive cores).
For the explanation, we only focus on extending all occurrences of~$C$ to the right to~$C C_s$ (the left side is done symmetrically)\JO{, cf.~\cref{figComputeLCE}}.
We maintain an array~$D$ of length~$\tau_P-1$ storing pairs~$(\Takai{X}{h}, \ell_h)$ with $\Takai{X}{h} \in \Takai{\Gamma}{h}$ and $\ell_h \ge 0$ for each height~$h \in [1..\tau_P-1]$ such that
$\pi(\Takai{X}{h})$ has the currently longest extension of length~$\ell_h$ with the core $\Takai{S}{h-1}$ of $P$ in common (cf.~\cref{eqCores}).
By maintaining~$D$, we can first query \fnLCE{} with the specific non-terminal in $D$, and then resort to plain symbol comparison.
We descend to the child where the mismatch happens and recurse until reaching the character level of \TreeT{}.
This all works since by the core property the mismatch of a child means that there is a mismatch in the expansion of this child.
Since a plain symbol comparison with matching symbols lets us exchange the currently used non-terminal in $D$ with a longer matched prefix or a different non-terminal with larger expansion,
we can bound (a) the total number of naive symbol matches to \Oh{m} and (b) the total number of naive symbol mismatches and LCE queries to $\Oh{\OccC \tau_T}$.
This concludes the right extension of $C$.
By symmetry, we also perform LCE queries in reversed direction to match the cores $\Takai{A}{h}$ on the left side of $C$ in \cref{eqCores} for all $h \in [1..\tau_P-2]$ to extend $C$ to the left.

\JFigure{\begin{figure}
\begin{minipage}{0.5\linewidth}
\includegraphics[width=\linewidth]{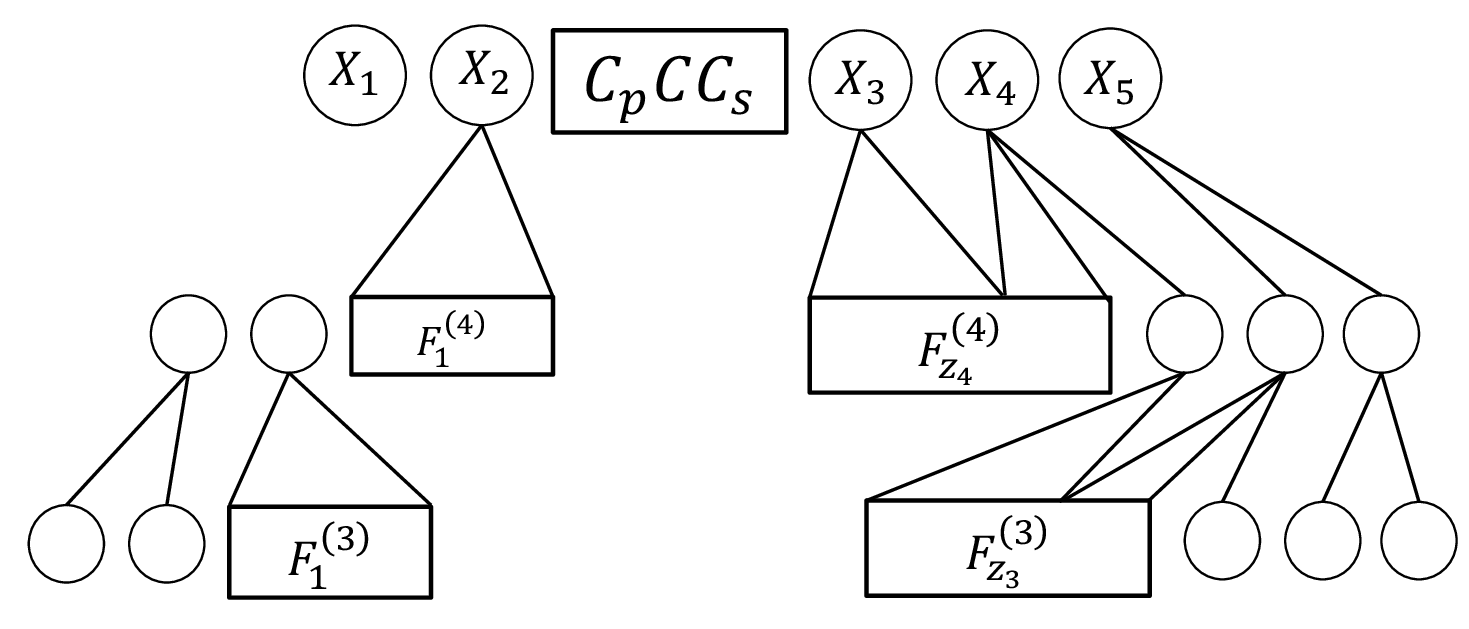}
\end{minipage}
\hfill
\begin{minipage}{0.5\linewidth}
  \caption{Occurrence of $C$ in \TreeT{}.
      Circles symbolize nodes, rectangles multiple sibling nodes.
      To check whether this occurrence corresponds to an occurrence of~$P$, it suffices to check the extensions of $X_1X_2C_pCC_sX_3X_4X_5$. 
      Although $X_1, X_2, X_3,X_4,X_5$ are not cores, $\pi(X_2)$, $\pi(X_3)$ and $\pi(X_4)$ extend to symbols that match with symbols in the derivation tree of~$\BunpouP$.
      For the pattern matching, it suffices to match $\Takai{F}{h}_{1}$ and two symbols to its left, and $\Takai{F}{h}_{z_h}$ and three symbols to its right (cf.~\cref{secCores}).
}
      \label{figExtendNonTerminals}
\end{minipage}
\end{figure}
}

\J{\section{LCE Queries in Detail}\label{secLCEDetail}
  \JO[We continue with the description in \cref{secExtendMatches} for the usage of LCE queries: We]{Put in technical terms, we} start with $D$ storing only invalid entries~$\bot$ to indicate that there is no candidate on any height for LCE queries.
Suppose that we are given a \DAG{} path $\Path := (\Takai{Y}{h^*},\ldots,\Takai{Y}{\tau_P})$ to an occurrence of~$C$ and an offset~$\Takai{o}{\tau_P}$ such that 
$\pi(\Takai{Y}{\tau_P})[\Takai{o}{\tau_P}..]$ starts with $C$.

In a procedure called \textsc{NextRightSymbol} in \cref{algExtendOccurrence},
we obtain the highest-positioned symbol 
immediately to the right of the occurrence of \(C\) along the given DAG path.
This skips levels where the expansion of \(C\) is a suffix of the current
non-terminal expansion. If the required symbol does not exist, the occurrence
is rejected; if the right extension is already complete, we proceed directly to
the final verification step.
Given this symbol is at height $j_0$,
we scan the entries of $D$ from $h = j_0$ to $h = 1$ while checking whether $D[h] = \bot$:

\begin{itemize}
  \item Suppose that $D[h] = (\Takai{\tilde{Y}}{h},  \Takai{\tilde{\ell}}{h})$ is not empty, 
    we compute $\ell := \fnLCE(\Takai{\tilde{Y}}{h}, 1, \Takai{Y}{h}, 1)$.
    We abort if $\ell < \Takai{\tilde{\ell}}{h}$.
    Otherwise, 
    we update $D[h] := (\Takai{Y}{h}, \ell)$, 
    set $\Takai{Y}{h-1} := \pi(\Takai{Y}{h})[\Takai{\tilde{\ell}}{h}+1]$, and recurse.
  
  \item If $D[h] = \bot$, we compare $\pi(\Takai{Y}{h})$ with the respective cores of the pattern at height~$h-1$, cf.~\cref{eqCores,figExtendNonTerminals}.
    Given we matched $\ell$ symbol pairs, we set $D[h] := (\Takai{Y}{h}, \ell)$.
\end{itemize}

On reaching $h = 1$, we compare the RHSs of $\Takai{Y}{1}$ and its right siblings with the remaining part of the pattern.
On reaching a mismatch or completing $\pi^*(C_s)$, we update the entries of~$D$ while climbing up \Path,
  and store the node~$\Takai{Y}{h^*}$ with the starting position of the respective occurrence of~$P$ relative to its grammar subtree in a list~$W$.
  The traversal, the update of~$D$, and LCE queries take \Oh{\tau_T} time per occurrence of~$C$,
  while the total number of naive symbol matches accumulate to \Oh{m} time.
  We need the path of $\lambda_L$ to start at $\Takai{Y}{h^*}$, since otherwise selecting sibling nodes on the same height can cost $\Oh{\tau_T}$ time. 
}\Cref{algLocateMain} and \cref{algExtendOccurrence} show pseudocode for the locate procedure described above. 
The algorithm consists of several key components: computing the core, extending occurrences using LCE queries, and finding the starting positions, which we describe next.

\begin{algorithm}
\begin{algorithmic}[1]
\Require Pattern $P$ of length $m$, Text $T$ with grammar $\BunpouT$
\Ensure List of starting positions of all occurrences of $P$ in $T$
\Statex \textbf{Step 1:} Compute grammar $\BunpouP$ of pattern $P$ and determine core $C$
\State Initialize: $h \gets 0$, $\Takai{P}{0} \gets P$
\While{LMS factorization of $\Takai{P}{h}$ has $> 2$ factors}
    \State Compute LMS factorization: $\Takai{P}{h} = \Takai{F}{h}_1 \cdots \Takai{F}{h}_{z_h}$
    \For{$i = 2$ \textbf{to} $z_h - 1$}
        \State $\Takai{X}{h+1}_i \gets \fnLookup(\Takai{F}{h}_i)$
        \If{$\Takai{X}{h+1}_i = \bot$}~\Return $\emptyset$ \Comment{pattern not in text}
        \EndIf
    \EndFor
    \State $\Takai{P}{h+1} \gets \Takai{X}{h+1}_2 \cdots \Takai{X}{h+1}_{z_h-1}$
    \State $h \gets h + 1$
\EndWhile
\State $\tau_P \gets h$
\Comment{Height of derivation tree of pattern grammar $\BunpouP$}
\State Determine core $C$ from $\Takai{P}{\tau_P}$ using rules in \cref{secCores}
\Statex \textbf{Step 2:} Find all occurrences of core $C$ in $\GST$
\State Compute locus $v$ of $C$ in $\GST$ using \fnChild{} queries
\State $L \gets \emptyset$
\For{each leaf $\lambda$ in subtree rooted at $v$}
    \State Extract $(\Takai{Y}{\tau_P}, \Takai{o}{\tau_P})$ from $\lambda$
    \For{each $\DAG$ path $(\Takai{Y}{h^*}, \ldots, \Takai{Y}{\tau_P})$ to lowest ancestor $\Takai{Y}{h^*}$ with $|P| \le |\pi^*(\Takai{Y}{h^*})|$}
      \State Add $((\Takai{Y}{h^*}, \ldots, \Takai{Y}{\tau_P}), \Takai{o}{\tau_P})$ to list $L$
      \Comment{Enumeration is over parent occurrences and can be implemented with a stack, cf.~Step 4.}
    \EndFor
\EndFor
\Statex \textbf{Step 3:} Extend core occurrences to pattern occurrences (if possible)
\State Initialize $D[1..\tau_P-1] \gets [\bot, \ldots, \bot]$ \Comment{Gets filled in \textsc{ExtendOccurrence}}
\State Initialize $W \gets \emptyset$
\For{each $((\Takai{Y}{h^*}, \ldots, \Takai{Y}{\tau_P}), \Takai{o}{\tau_P})$ in $L$}
\State \textsc{ExtendOccurrence}$((\Takai{Y}{h^*}, \ldots, \Takai{Y}{\tau_P}), \Takai{o}{\tau_P}, D, W)$ \Comment{cf.~\cref{algExtendOccurrence}}
\EndFor
\Statex \textbf{Step 4:} Compute starting positions in $T$
\State Initialize $\text{result} \gets \emptyset$
\For{each $(X, \ell)$ in $W$}
  \State Initialize visit stack $\mathcal{V} \gets [(X,\ell)]$
  \While{$\mathcal{V}$ is not empty}
      \State $(\text{curr}, \text{pos}) \gets \Call{pop}{\mathcal{V}}$
      \If{$\text{curr}$ is the root of $\DAG$}
          \State Add $\text{pos}$ to $\text{result}$
      \Else
          \For{each parent occurrence $(Y,i)$ of $\text{curr}$ in $\DAG$}
              \Comment{$\pi(Y)[i]=\text{curr}$}
              \State $\text{pos}' \gets \text{pos} + Y.L[i-1]$
              \Comment{$Y.L[i-1]=\sum_{j=1}^{i-1}|\pi^*(\pi(Y)[j])|$ and $Y.L[0]=0$}
              \State \Call{push}{$\mathcal{V},(Y,\text{pos}')$}
          \EndFor
      \EndIf
  \EndWhile
\EndFor
\State \Return $\text{result}$
\end{algorithmic}
\caption{Locate: Finding all occurrences of $P$ in $T$.}
\label{algLocateMain}
\end{algorithm}

\begin{algorithm}
\begin{algorithmic}[1]
\Require $\DAG$ path $(\Takai{Y}{h^*}, \ldots, \Takai{Y}{\tau_P})$, offset $\Takai{o}{\tau_P}$, array $D$, result list $W$
\Ensure Updates $W$ with valid pattern occurrences
\State $(j_0, Y^{(j_0)}) \gets \Call{NextRightSymbol}{(Y^{(h^*)},\ldots,Y^{(\tau_P)}), o^{(\tau_P)}, C}$
\If{$(j_0,Y^{(j_0)})=\bot$}
    \State \textbf{goto} \textsc{FinalVerification}
\EndIf
\For{$j = j_0$ \textbf{down to} $1$}
    \If{$D[j] \neq \bot$}
        \State Let $D[j] = (\Takai{\tilde{Y}}{j}, \Takai{\tilde{\ell}}{j})$ \Comment{$\Takai{\tilde{\ell}}{j}$ is the length of the longest common prefix so far}
        \State $\ell \gets \fnLCE(\Takai{\tilde{Y}}{j}, 1, \Takai{Y}{j}, 1)$
        \If{$\ell < \Takai{\tilde{\ell}}{j}$}~\Return \Comment{abort this occurrence}
        \Else
            \State $\Takai{Y}{j-1} \gets \pi(\Takai{Y}{j})[\Takai{\tilde{\ell}}{j} + 1]$
            \State \textbf{continue} to next iteration
        \EndIf
    \ElsIf{$D[j] = \bot$}
        \State $\Takai{\ell}{j} \gets $ longest common prefix of $\pi(\Takai{Y}{j})$ and pattern cores at height $j-1$
        \State Update $D[j] \gets (\Takai{Y}{j}, \Takai{\ell}{j})$
        \State $\Takai{Y}{j-1} \gets \pi(\Takai{Y}{j})[\Takai{\ell}{j} + 1]$
        \State \textbf{continue} to next iteration
    \EndIf
\EndFor
\Statex \textsc{FinalVerification:}
\State Compare $\pi(\Takai{Y}{1})$ and right siblings with remaining pattern
\If{complete match found}
    \State Compute starting position $\ell$ relative to $\Takai{Y}{h^*}$ from the input $\DAG$ path
\State Add $(\Takai{Y}{h^*}, \ell)$ to $W$
\EndIf
\State Update $D$ while climbing back up the path
\end{algorithmic}
\caption{Extend a core occurrence to check if it corresponds to a pattern occurrence.
\Call{NextRightSymbol}{} returns the highest-positioned symbol immediately to the right of the occurrence of \(C\) along the given DAG path, skipping levels where the expansion \(C\) is a suffix of the expansion of the respective non-terminal; it returns $\bot$ if the right extension is already complete; if the required symbol is not available, the occurrence is rejected.
}
\label{algExtendOccurrence}
\end{algorithm}

\paragraph{Finding the Starting Positions}
It is left to compute the starting position in~$T$ of each occurrence captured by an element in $W$.
We can do this similarly to computing the pre-order ranks in a tree:
For each pair~$(X,\ell) \in W$, climb up \DAG{} from~$X$ to the root while accumulating the expansion lengths of all left siblings of the nodes we visit 
(we can make use of $Y.L$ for $Y$ being a parent of $X$).
If this accumulated length is $s$, then $\ell+s$ is the starting position of the occurrence captured by $(X,\ell)$.
However, this approach would cost \Oh{\tau_T} time per element of~$W$.
Here, we use the amortization argument of \citet[Sect.~5.2]{claude12grammarindex},
which works if we augment, in a pre-computation step, each non-terminal~$X$ in $\Gamma$ with 
(a) a pointer to the lowest ancestor~$Y_X$ on every path from~$X$ to the \DAG{} root that has~$X$ at least twice as a descendant, 
and
(b) the lengths of the expansions of the left siblings of the child of~$Y_X$ being a parent of~$X$ or $X$ itself.
By doing so, when taking a pointer of a non-terminal~$X$ to its ancestor~$Y_X$, we know that~$X$ has another occurrence in \DAG{} (and thus there is another occurrence of~$P$). 
Therefore, we can charge the cost of climbing up the tree with the amount of occurrences~\Occ{} of the pattern.

\paragraph{Total Time}
To sum up, 
we need \Oh{m \lg |\Symbols|} time for finding~$C$,
\Oh{\OccC \tau_T \lg |\Symbols|} time for computing the non-terminals covering~$C$,
\Oh{m + \OccC \tau_T} time for reducing these non-terminals to~$W$, 
and \Oh{\Occ} time for retrieving the starting positions of the occurrences of~$P$ in~$T$ from~$W$.
To be within our \Oh{g} space bounds, we can process each \DAG{} parent of~$C$ individually, and keep only~$D$ globally stored during the whole process.
The total additional space is therefore $\Oh{\tau_T} \subset \Oh{g}$ for maintaining~$D$ and a path for each occurrence of~$C$.
So we finally obtain the claim of \cref{thmLocatePattern}.

\section{Walk-through Example}\label{secExample}

To illustrate the GCIS index construction and pattern matching algorithm, 
we present a detailed example using a binary alphabet $\Sigma = \{\texttt{a}, \texttt{b}\}$ with $\texttt{a} < \texttt{b}$.

\paragraph{Text and Pattern}
Consider the text $T = \texttt{abaababaabaab}$ of length $n = 13$ and pattern $P = \texttt{abaab}$ of length $m = 5$.
The pattern occurs 3 times in the text at positions 1, 6, and 9.

\paragraph{Building the GCIS Grammar of the Text}
We first construct the GCIS grammar $\BunpouT$ for the text $T$.

\textbf{Height 0:} $\Takai{T}{0} = \texttt{abaababaabaab}$

We identify the suffix types:
\begin{itemize}
\item Positions with type \typeS: $\{1, 3, 4, 6, 8, 9, 11, 12, 13\}$
\item Positions with type \typeL: $\{2, 5, 7, 10\}$
\item Positions with type \typeHoshi{} (LMS): $\{1, 3, 6, 8, 11\}$
\end{itemize}

The LMS factorization gives us:
$$\texttt{\#}T = \texttt{\#ab} \cdot \texttt{aab} \cdot \texttt{ab} \cdot \texttt{aab} \cdot \texttt{aab}$$

After omitting the delimiter \texttt{\#}, we obtain factors:
\begin{align*}
\Takai{F}{0}_1 &= \texttt{ab} \\
\Takai{F}{0}_2 &= \texttt{aab} \\
\Takai{F}{0}_3 &= \texttt{ab} \\
\Takai{F}{0}_4 &= \texttt{aab} \\
\Takai{F}{0}_5 &= \texttt{aab}
\end{align*}

We assign non-terminals 
$\Takai{X}{1}_1 \rightarrow \texttt{ab}$ and
$\Takai{X}{1}_2 \rightarrow \texttt{aab}$ 
with $\Takai{X}{1}_1 \succ \Takai{X}{1}_2$ based on the lexicographic order of their expansions.

\textbf{Height 1:} $\Takai{T}{1} = \Takai{X}{1}_1 \Takai{X}{1}_2 \Takai{X}{1}_1 \Takai{X}{1}_2 \Takai{X}{1}_2$

The LMS factorization yields:
$$\Takai{T}{1} = (\Takai{X}{1}_1) \cdot (\Takai{X}{1}_2 \Takai{X}{1}_1 \Takai{X}{1}_2 \Takai{X}{1}_2)$$

We create: $\Takai{X}{2}_1 \rightarrow \Takai{X}{1}_1$ and $\Takai{X}{2}_2 \rightarrow \Takai{X}{1}_2 \Takai{X}{1}_1 \Takai{X}{1}_2 \Takai{X}{1}_2$

Thus: $\Takai{T}{2} = \Takai{X}{2}_1 \Takai{X}{2}_2$

Since the factorization has only 2 factors, we stop here with $\tau_T = 2$ and start symbol $\Takai{X}{3} \rightarrow \Takai{X}{2}_1 \Takai{X}{2}_2$.

The complete grammar $\BunpouT$ consists of:
\begin{align*}
\Takai{X}{3} &\rightarrow \Takai{X}{2}_1 \Takai{X}{2}_2 \\
\Takai{X}{2}_1 &\rightarrow \Takai{X}{1}_1 \\
\Takai{X}{2}_2 &\rightarrow \Takai{X}{1}_2  \Takai{X}{1}_1 \Takai{X}{1}_2 \Takai{X}{1}_2 \\
\Takai{X}{1}_1 &\rightarrow \texttt{ab} \\
\Takai{X}{1}_2 &\rightarrow \texttt{aab}
\end{align*}

with grammar size $g = 2 + 1 + 4 + 2 + 3 = 12$.

\paragraph{Building the GCIS Grammar of the Pattern}
Now we construct $\BunpouP$ for pattern $P = \texttt{abaab}$.

\textbf{Height 0:} $\Takai{P}{0} = \texttt{abaab}$

LMS positions are $\{1, 3\}$, giving factorization:
$$\texttt{\#}P = \texttt{\#ab} \cdot \texttt{aab}$$

We check $\fnLookup(\texttt{ab}) = \Takai{X}{1}_1$ and $\fnLookup(\texttt{aab}) = \Takai{X}{1}_2$. 
Both exist in $\BunpouT$.
Thus, we have $\Takai{P}{1} = \Takai{X}{1}_1 \Takai{X}{1}_2$.
Since we are left with two symbols,
we already stop here with $C_p = \Takai{X}{1}_1$, $C = \Takai{X}{1}_2$.

\paragraph{Finding and Extending Core Occurrences}
$C$ appears three times in the RHS of $\Takai{X}{2}_2$, so the number of core occurrences $\OccC$ is three.
For each core occurrence, we check if it can be extended to the full pattern $P = \pi^*(\Takai{X}{1}_1 \Takai{X}{1}_2)$:
The second and third occurrence of $C$ in $\Takai{X}{2}_2$ have a sufficient distance from the beginning of the expansion of $\Takai{X}{2}_2$ to accommodate the full pattern,
so we perform reversed-LCE queries on them to find the longest common suffix with $C_p = \texttt{ab}$.
We thus obtain the second occurrence of $P$ at position 6 and the third occurrence of $P$ at position 9.
For the first occurrence, we need to check each parent of $\Takai{X}{2}_2$ to see if it can accommodate the full pattern.
There is only one parent $\Takai{X}{3}$ of $\Takai{X}{2}_2$, and the left sibling of $\Takai{X}{2}_2$ in $\Takai{X}{3}$ is $\Takai{X}{2}_1$, on which we apply reversed-LCE queries to check if it can match~$C_p$.
This gives us the first occurrence of $P$ at position 1 in the expansion of $\Takai{X}{3}$.

\section{Implementation}\label{secPractice}
The implementation deviates from theory with respect to the rather large hidden constant factor in the \Oh{g} words of space.
We drop \GST{}, and represent \DAG{} with multiple arrays.
For that, we first enumerate the non-terminals as follows:
The height and the lexicographic order induce a natural order on the non-terminals in $\Gamma$, 
which are ranked by first their height and secondly by the lexicographic order of their RHSs,
such that we can represent $\Gamma = \{X_1, \ldots, X_{|\Gamma|}\}$.
By stipulating that all characters are lexicographically smaller than all non-terminals,
we obtain the property that $\pi(X_i) \prec \pi(X_{i+1})$ for all $i \in [1..|\Gamma|-1]$.
In the following, we first present a plain representation of $\DAG$, called \GCISnep{}, 
then give our modified \fnLocate{} algorithm, and subsequently present a compressed version of \DAG{} using universal coding, called \GCISuni{}.
Finally, we evaluate both implementations in \cref{secExperiments}.

Our first implementation, called \GCISnep{}\footnote{\GCISnep{} stands for GCIS with \textbf{n}on-terminals \textbf{e}ncoded \textbf{p}lainly.},
represents each symbol with a 32-bit integer.
We use $R := \prod_{i=1}^{|\Gamma|} \pi(X_i)$ 
again, but omit the delimiters \texttt{\$} separating the RHSs.
To find the RHS of a non-terminal $X_i$, 
we create an array of positions~$Q[1..|\Gamma|]$ such that $Q[i]$ points to the starting position of $\pi(X_i)$ in $R$.
Finally, we create an array $L[1..|\Gamma|]$ storing the length of the expansion $|\pi^*(X_i)|$ in $L[i]$, for each non-terminal~$X_i$. 
Due to the stipulated order of the symbols, the strings $R[Q[i]..Q[i+1]-1]$ are sorted in ascending order.
Hence, we can evaluate $\fnLookup(S)$ for a string~$S$ in $\Oh{|S| \lg |\Symbols|}$ time by a binary search on $Q$ with $i \mapsto R[Q[i]..Q[i+1]-1]$ as keys.

\paragraph{Locate}
Our implementation follows theory for computing \BunpouP{} and $C$ (cf.~\cref{secCores}) in the same time bounds, but deviates after computing the core~$C$:
To find all non-terminals whose RHSs contain $C$, 
we linearly scan the RHSs of all non-terminals on height~$\tau_P$, 
which we can do cache-friendly since the RHSs of $R$ are sorted by the height of their respective non-terminals. 
This takes \Oh{g + |C|} time in total with a pattern matching algorithm~\cite{knuth77kmp}.
Finally, for extending a found occurrence of the core~$C$ to an occurrence of~$P$, 
we follow the naive approach to descend \DAG{} to the character level and compare the expansion with $P$ character-wise, which results in \Oh{\OccC{} |P| \tau_T} time.
The total time cost is $\Oh{g + |P| (\OccC{} \tau_T + \lg |\Symbols|)}$.

\JFigure{\begin{figure}
\begin{minipage}{0.55\linewidth}
\includegraphics[width=\linewidth]{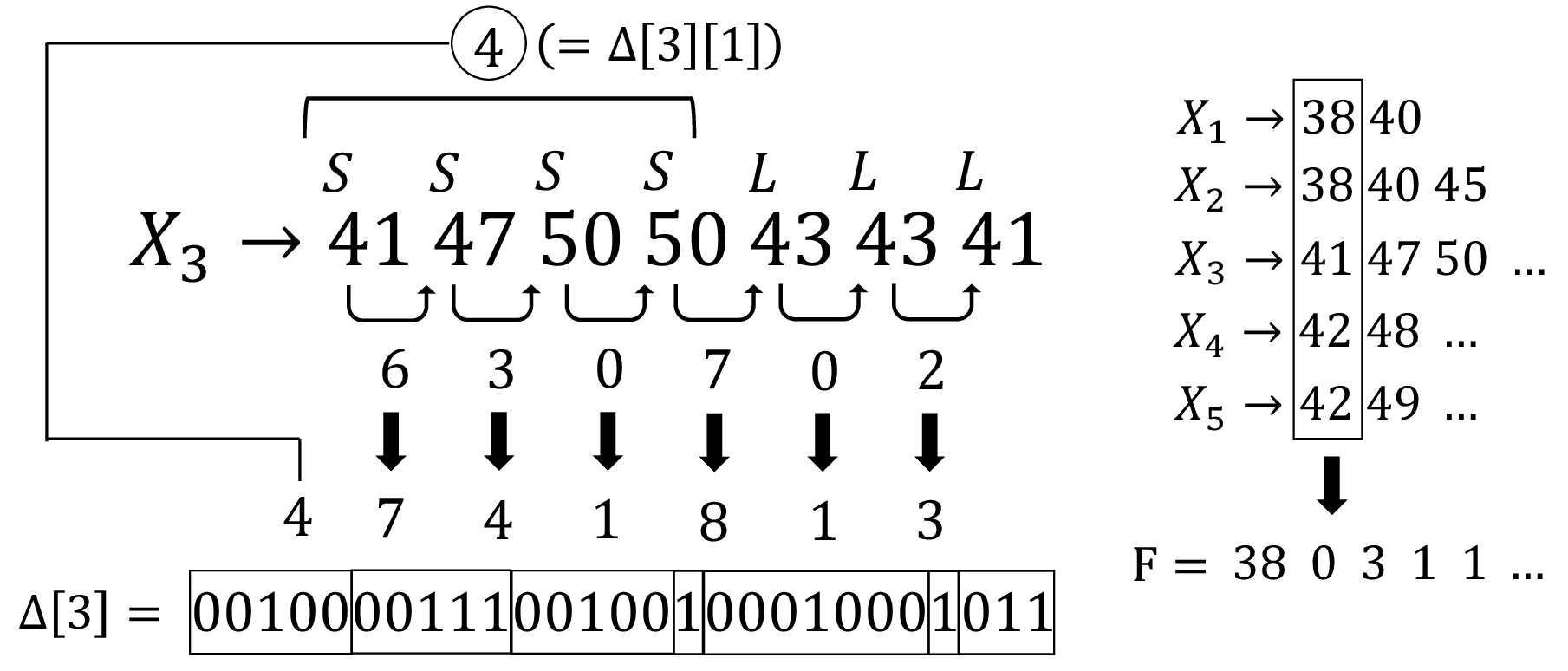}
\end{minipage}
\hfill
\begin{minipage}{0.4\linewidth}
      \caption{Encoding of \GCISuni{}.
        \emph{Right:} Encoding the first symbol of each RHS with Elias--Fano and storing the sequence in an integer array~$F$.
        \emph{Left:} Storing the remaining characters of $\pi(X_3)$ in $\Delta[3]$ delta-encoded with Elias-$\gamma$. $\pi(X_3)$ is a bitonic sequence, where the first $\Delta[3][1] = 4$ symbols are increasing. 
      The bottom row depicts the binary representation of $\Delta[3]$.
}
      \label{figGCISuni}
\end{minipage}
\end{figure}
}

\begin{table}[t]
\caption{Top: Sizes of the used datasets and the indexes stored on disk. 
Sizes are in megabytes [MB]. 
Bottom: Statistics of the GCIS grammar.
}
\label{tabFilesize}
\hspace{0.5em}
\begin{tabular}{|l|r||r|r|r|r|r|}
\hline
dataset & input size & \GCISnep{} & \GCISuni{} & ESP-index & FM-index & r-index  \\ \hline
\textsc{commoncrawl    } & 221.180 & 220.119 & 138.856 & 156.006 & 122.575 & 454.124 \\
\textsc{dna            } & 403.927 & 527.553 & 327.852 & 297.001 & 216.153 & 2123.817 \\
\textsc{einstein.de} & 92.758  & 1.139   & 0.428   & 0.697   & 40.291  & 1.146 \\
\textsc{english.001.2  } & 104.857 & 14.784  & 7.489   & 10.464  & 46.981  & 14.389 \\
\textsc{fib41          } & 267.914 & 0.001   & 0.001   & 0.001   & 71.305  & 0.007 \\
\textsc{influenza      } & 154.808 & 23.373  & 13.871  & 15.729  & 53.066  & 28.775 \\
\textsc{kernel         } & 257.961 & 21.298  & 10.469  & 12.545  & 125.087 & 28.947 \\
\textsc{rs.13          } & 216.747 & 0.002   & 0.001   & 0.002   & 57.653  & 0.009 \\
\textsc{tm29           } & 268.435 & 0.002   & 0.001   & 0.002   & 69.347  & 0.009 \\
\textsc{world leaders  } & 46.968  & 5.415   & 2.573   & 3.611   & 21.097  & 5.627 \\
\hline
\end{tabular}

\begin{tabular}{|l|r|r|r|} \hline
dataset & $|\Gamma|$ & $g$ & $|\pi(\Takai{X}{\tau_T})|$ \\
\hline
\textsc{commoncrawl  }& 8565089  & 37899804 & 7603482 \\
\textsc{dna          }& 21068903 & 89750482 & 11462490\\
\textsc{einstein.de  }& 50554    & 183664   & 3491\\
\textsc{english.001.2}& 661175   & 2373818  & 152854\\
\textsc{fib41        }& 67       & 173      & 22 \\
\textsc{influenza    }& 971626   & 3900153  & 477072 \\
\textsc{kernel       }& 943991   & 3436501  & 2478 \\
\textsc{tm29         }& 104      & 311      & 16 \\
\textsc{world leaders}& 246589   & 860611   & 31743\\
\hline
\end{tabular}
\end{table}

\paragraph{\GCISuni{}}
To save space, we can leverage universal code to compress the RHSs of the productions.
First, we observe that $Q$ and the first symbols $F := \pi(X_1)[1], \ldots, \pi(X_{|\Gamma|})[1]$ form an ascending sequence, 
such that we represent both $Q$ and $F$ in Elias--Fano coding~\cite{elias74code}.
Next, we observe that each RHS $\pi(X_i)$ forms a bitonic sequence: 
the ranks of the first $\ell_i$ symbols are non-decreasing, while the rest of the ranks are non-increasing. 
Our idea is to store $\ell_i$ and the rest of $\pi(X_i)[2..]$ in delta-coding, i.e., $\Delta[i][k] := |\pi(X_i)[k] - \pi(X_i)[k-1]|$ for $k \in [2..|\pi(X_i)|]$,
which is stored in Elias-$\gamma$ code~\cite{elias75universal}.
Although $\pi(X_i)[k] - \pi(X_i)[k-1] < 0$ for $k > \ell_i$,
we can decode $\pi(X_i)[k]$ by subtracting instead of adding the difference to $\pi(X_i)[k-1]$ as usual in delta-coding.
Hence, we can replace $R$ with $\Delta$, but need to adjust $Q$ such that $Q[i]$ points to the first bit of $\Delta[i]$.
Finally, like in the first variant, we store the expansion lengths of all non-terminals in $L$.
Here, we separate $L$ in a first part using 8 bits per entry, then 16 bits per entry, and finally 32 bits per entry. 
To this end, we represent $L$ by three arrays, 
start with filling the first array, 
and continue with filling the next array 
whenever we process a value whose bit representation cannot be stored in a single entry of the current array.
Since Elias--Fano code supports constant-time random access and Elias-$\gamma$ supports constant time per decoded codeword during sequential access,
we can decode $\pi(X_i)$ by accessing~$F[i]$ and then sequentially decode $\Delta[i]$.
Hence, we can simulate \GCISnep{} with this compressed version without sacrificing the theoretical bounds.
We call the resulting index \GCISuni{}.
\JO{\Cref{figGCISuni} captures our compression steps.}

\section{Experiments}\label{secExperiments}
In the following we present an evaluation of our \CPlusPlus{} implementation and different self-indexes for comparison, which are
the FM-index~\cite{ferragina08compressed},
the ESP-Index~\cite{takabatake14esp}, and
the r-index~\cite{gagie18bwt}\footnote{See \url{https://github.com/mpetri/FM-Index}, \url{https://github.com/tkbtkysms/esp-index-I}, and \url{https://github.com/nicolaprezza/r-index}, respectively.}.
All code has been compiled with \texttt{gcc-10.2.0} in the highest optimization mode \texttt{-O3}.
We ran all our experiments \JO{on a Mac Pro Server} with an Intel Xeon CPU X5670 clocked at 2.93GHz running Arch Linux.

\begin{figure}[t]
\vspace{-1em}
  \includegraphics[width=\linewidth]{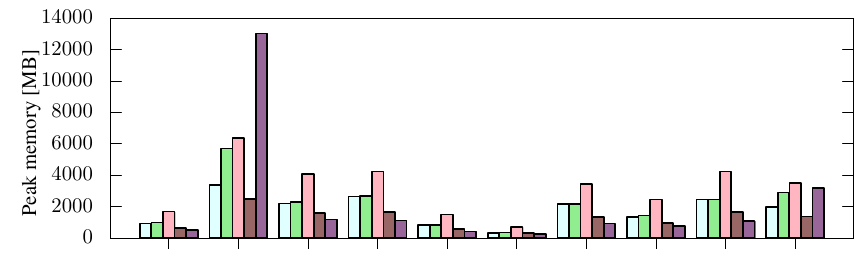}
\vspace{-1em}
\includegraphics[width=\linewidth]{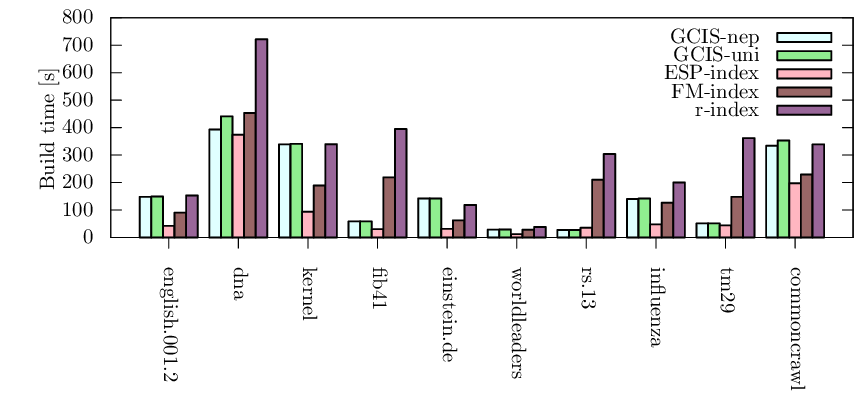}
\caption{Maximum memory consumption \emph{(top)} and time \emph{(bottom)} during the construction of the indexes. 
      }
      \label{figIndexConstruction}
\vspace{-1em}
\end{figure}

Our datasets shown in \cref{tabFilesize} are from the Pizza\&Chili and the tudocomp~\cite{dinklage17tudocomp} corpus.\footnote{To save space, we renamed the datasets \textsc{commoncrawl.ascii.txt} and \textsc{einstein.de.txt} to \textsc{commoncrawl} and \textsc{einstein.de}, respectively.}
With respect to the index sizes, we empirically observe the ranking \GCISuni{} $<$ ESP-index $<$ \GCISnep{}, followed by one of the BWT-based indexes.
While the $r$-index needs less space than the FM-index on highly-compressible datasets, it is the least favorable option of all indexes for less-compressible datasets.
\Cref{figIndexConstruction} gives the time and memory needed for constructing the indexes.
\JO{There, we measured the memory as the \emph{maximum resident set size} reported by \texttt{/usr/bin/time}.}

The bottom of \cref{tabFilesize} presents some characteristics of the GCIS grammar, where $|\pi(\Takai{X}{\tau_T})|$ is the size of the start symbol.
We can see that the number of non-terminals, grammar size~$g$, and the size of the start symbol scale with the compressibility of the input.
Especially artificially generated datasets compress well with GCIS\@.

\begin{figure}
\begin{minipage}{0.5\linewidth}
\includegraphics[width=\linewidth]{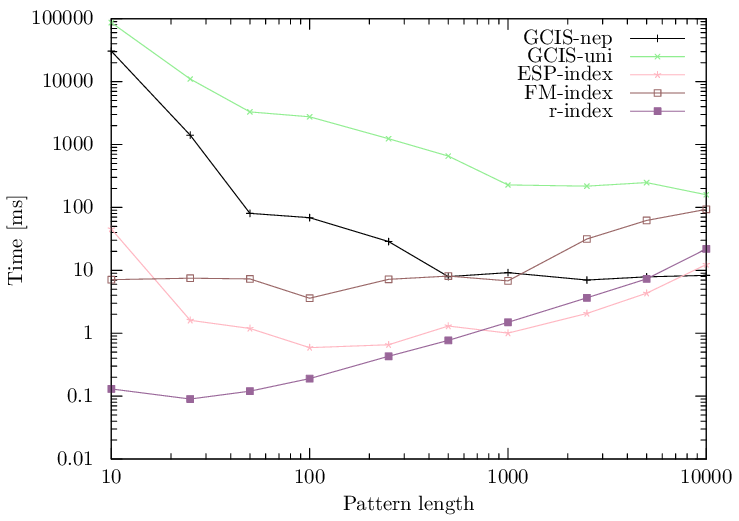}
\end{minipage}
\begin{minipage}{0.5\linewidth}
\includegraphics[width=\linewidth]{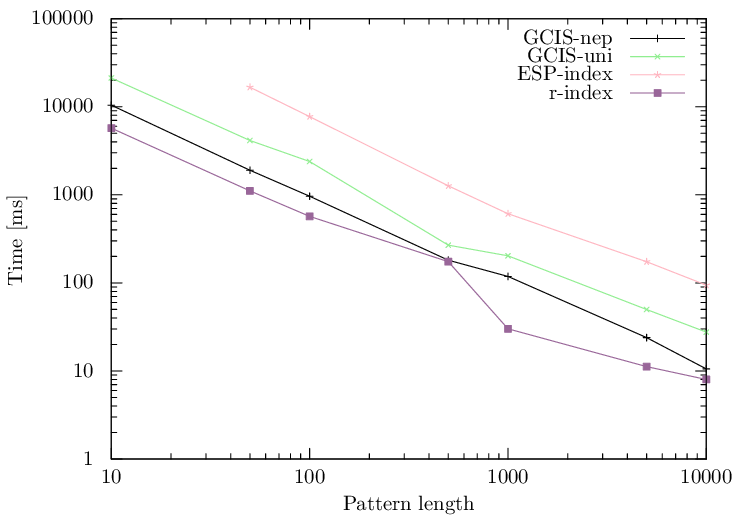}
\end{minipage}
      \caption{Time for \protect\fnLocate{} while scaling the pattern length 
      on the datasets \textsc{english.001.2} (left) and \textsc{fib41} (right).
      The plots are in logscale.
      The right figure does not feature the FM-index, which takes considerably more time than the other approaches.
      For the same reason, there is no data shown for the ESP-index for small pattern lengths, which needs 170 seconds on average for $|P| = 10$.
      }
      \label{figTimeLocate}
\vspace{-1em}
\end{figure}

\begin{figure}
\begin{minipage}{0.5\linewidth}
\includegraphics[width=\linewidth]{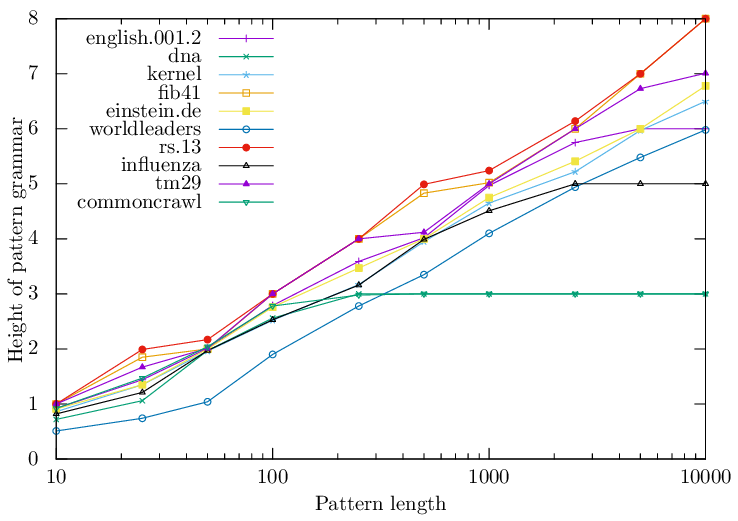}
\end{minipage}
\begin{minipage}{0.5\linewidth}
\includegraphics[width=\linewidth]{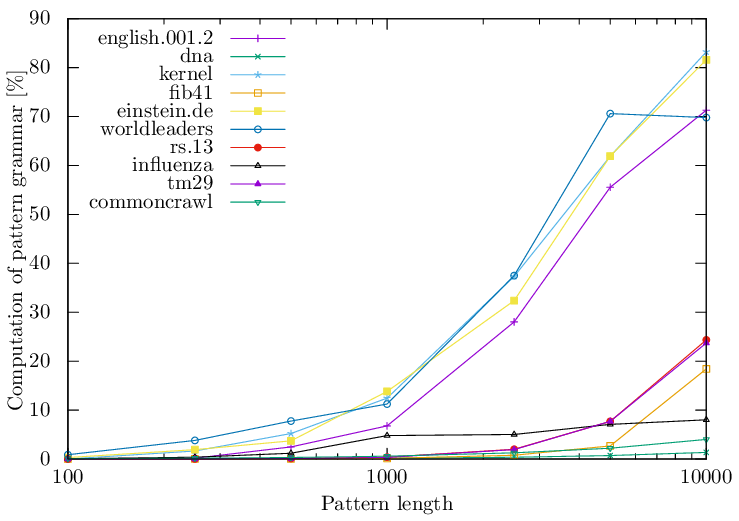}
\end{minipage}
\caption{\emph{Left:} The average height~$\tau_P$ of \BunpouP{} for a pattern of a certain length.
  \emph{Right:} Percentage of the computation of \BunpouP{} in relation to the whole running time for answering $\protect\fnLocate(P)$ with \GCISnep{}.
}
      \label{figPatternHeightPercentage}
\vspace{-2em}
\end{figure}

We can observe in \cref{figTimeLocate} that
our indexes answer $\fnLocate(P)$ fast when~$P$ is sufficiently long or has many occurrences \Occ{} in~$T$.
\GCISuni{} is always slower than \GCISnep{} due to the extra costs for decoding.
In particular for \textsc{english.001.2}, 
\GCISnep{} is the fastest index when the pattern length reaches  10000 characters and more. 
At this pattern length, the pattern grammar reached a height~$\tau_P$ of almost six, which is the height~$\tau_T$. The algorithm can extend an occurrence of a core to a pattern occurrence by checking only 80--100 characters.
However, when the pattern surpasses 5000 characters, 
the computation of \BunpouP{} becomes the time bottleneck.
In that respect, the ESP-index shares the same characteristic.
The universal encoding slows down locate time by roughly a factor of 2 to 10.
Let us have a look at the dataset \textsc{fib41},
which is linearly recurrent~\cite{du14decision}, 
a property from which we can derive the fact that a pattern that occurs at least once in~$T$ has actually a huge number of occurrences in~$T$.
There are almost 3,000,000 occurrences for patterns of length 100 in total.
Here, we observe that our indexes are faster than ESP-index.
ESP-index needs more time for \fnLocate{} than GCIS 
because GCIS can form a core that covers a higher percentage of the pattern than the core selected by ESP.
FM-index, and ESP-index with $|P| = 10$ take 100 seconds or more on average  -- we omitted them in the graph to keep the visualization clear.

In \cref{figPatternHeightPercentage}, 
we study the average pattern grammar height $\tau_P = \Oh{\lg |P|}$.
For this experiment, we created for each data point 100 patterns by randomly selecting a position~$j$ in~$T$ from which we extract a substring of the given length as a pattern.
For every dataset, we could observe that $\tau_P$ is logarithmic in the pattern length,
especially for the artificial datasets \textsc{fib41}, \textsc{tm29}, and \textsc{rs.13}, where $\tau_P$ is empirically larger than measured in other datasets.
In \textsc{dna} and \textsc{commoncrawl}, $\tau_P$ is at most 3, but this is because $\tau_T = 3$ for these datasets.

\begin{figure}
\begin{minipage}{0.5\linewidth}
\includegraphics[width=\linewidth]{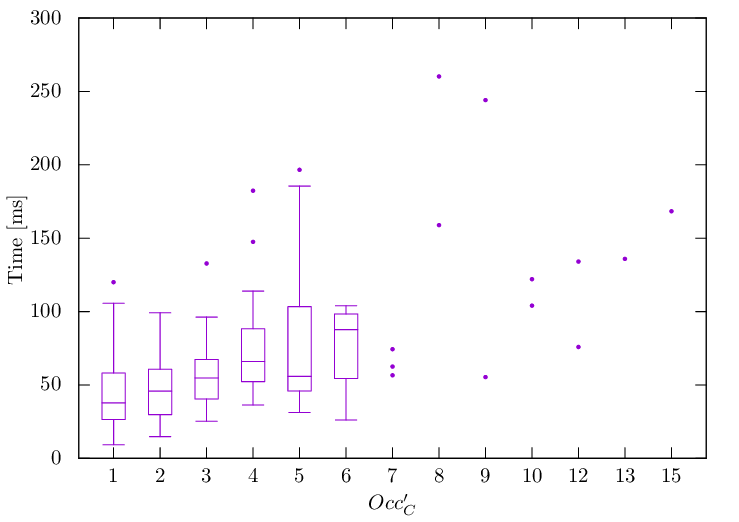}
\end{minipage}
\begin{minipage}{0.5\linewidth}
\includegraphics[width=\linewidth]{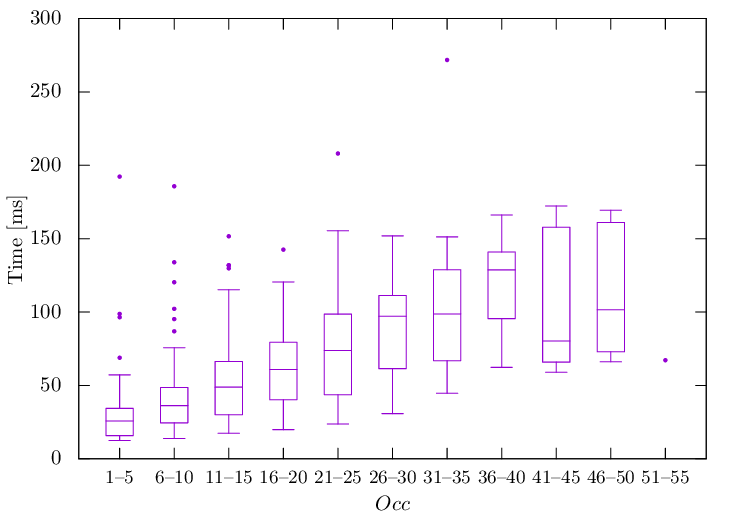}
\end{minipage}
      \caption{Relation of the number of visited \DAG{} nodes~$\OccC{}'$ (left)
      or the number of occurrences of the pattern~\Occ{} (right)
      with the time needed by \GCISnep{} for answering $\protect\fnLocate{}(P)$ on the dataset \textsc{english.001.2} for patterns of fixed length $|P| = 100$.
      }
      \label{figRelationOccTime}
\end{figure}

Next, we investigate the influence of the occurrences on the query time of \fnLocate{} on our indexes in \cref{figRelationOccTime}.
Instead of $\OccC{}$, which is almost always less than 5 for \textsc{english.001.2}, we measure $\OccC{}' \ge \OccC{}$ with $\OccC{}' = \Oh{\OccC{} \lg n}$ being the number of all visited \DAG{} nodes for finding the lowest ancestors of~$C$ whose expansion is large enough to extend that occurrence of $C$ to an occurrence of $P$.
For that, we focus again on \textsc{english.001.2} for patterns of fixed length $|P| = 100$, 
where we observe that $\OccC{}'$ has a stronger influence on the running time than \Occ{} has.

As a practical optimization, we prematurely abort the computation of \BunpouT{} at a certain height~$\tau' < \tau_T$
when the reduction of $\Takai{T}{\tau'-1}$ to $\Takai{T}{\tau'}$ with the newly introduced rules would increase the grammar size (we measure the needed space in terms of \GCISnep{}).
Hence, we may end up with a larger RHS of the start symbol $\Takai{X}{\tau'} \rightarrow \Takai{T}{\tau'-1}$,
which could be in fact the text itself if the text is incompressible with GCIS\@.
This heuristic takes effect in non-highly-repetitive datasets such as 
\textsc{dna} and \textsc{commoncrawl}. 
It also has an effect on the height $\tau_P$ since the core must have a height less than the height of the start symbol of~$T$.
Without this heuristic, we reach a height $\tau_T = 9$ for \textsc{dna},
and we also reach high values for $\tau_P$, on average $\tau_P = 6.55$
for $|P| = 10000$, or $\tau_P = 5.81$ for $|P| = 5000$.

Finally, the parameter $\OccC$ (core occurrences in grammar rules) is crucial for query performance.
We expect small values of $\OccC$ when
the core corresponds to high-level non-terminals (large height $h$) and is used in few rules.
As a rule of thumb we expect small $\OccC$ values for 
large pattern lengths ($m \ge 1000$) on highly repetitive texts ($g < 0.1n$)
when the pattern occurs in the text and the pattern grammar depth $\tau_P$ approaches the text grammar depth $\tau_T$.
In fact, under these conditions, GCIS index performs best, often outperforming alternatives by exploiting the small $\OccC$ to avoid excessive verification work.

\section{Practical Observations and Analysis}\label{secPracticalAnalysis}

We recall that our practical implementations \GCISnep{} and \GCISuni{} deviate from the theoretical algorithm in several ways:

\begin{itemize}
\item Premature stopping of grammar construction based on size heuristic, obtaining a height~$\tau' < \tau_T$.
\item Skipping the use of the generalized suffix tree GST altogether.
\item Binary search on sorted RHSs for $\fnLookup$.
\item Linear scanning for finding core occurrences
\item Naive character-by-character comparison instead of LCE: 
  While the optimized LCE-based extension takes amortized $\Oh{m + \OccC \tau_T}$ time, we use naive character-by-character comparison taking $\Oh{\OccC m \tau_T}$ time.
\end{itemize}

\subsection{Shorter Grammar Heights}
For the first bullet point, we terminate grammar construction at height $\tau' < \tau_T$ if continuing would increase the grammar size as measured in the space metric of the \GCISnep{} representation.
As a rule of thumb, one could define a threshold $t \in (0,1]$ and a weight~$w_0 \ge 0$ (both real non-negative numbers) that give a criterion 
for stopping the grammar construction at height $h$ when the compressed representation becomes larger than the previous level.
\[
\frac{|\Takai{T}{h-1}|}{|\Takai{T}{h}| + w_0 |\Takai{\Gamma}{h}|} < t.
\]
The rationale is that the heuristic prevents pathological cases where recursive factorization creates many small non-terminals with high overhead.
For incompressible or weakly compressible texts, forcing deeper factorization can expand $g$ due to creating unique factors and thus non-repeated non-terminals.
These non-terminals add additional overhead since we also need to keep track of their metadata (pointers, arrays) in the index.
The practical impact of this heuristic differs based on dataset characteristics:
\begin{itemize}
\item \textsc{dna}, \textsc{commoncrawl}: Natural stopping at $\tau_T = 3$ due to low repetition
\item \textsc{english.001.2}: Stops at $\tau_T = 6$ balancing compression and overhead
\item \textsc{fib41}, \textsc{tm29}, \textsc{rs.13}: Deep recursion ($\tau_T = 20+$) provides exponential compression
\end{itemize}
For datasets where the heuristic stops early (small $\tau_T$), patterns cannot achieve deep factorization, limiting the effectiveness of a core.
However, as shown in \cref{figPatternHeightPercentage}, even with $\tau_T = 3$ on \textsc{dna}, patterns reach $\tau_P = 2$--$3$ for $m \ge 1000$.

\subsection{Performance Relative to Theory}
On the one hand,
the practical implementation tracks theoretical behavior in the following cases: 

\begin{description}
  \item[Long patterns with high compression] For $m \ge 1000$ on highly repetitive texts (like \textsc{fib41}, \textsc{tm29}), the pattern grammar depth $\tau_P$ approaches the text grammar depth $\tau_T$.
The core covers a large portion of the pattern, making $\OccC$ small.
As shown in \cref{figTimeLocate}, GCIS becomes faster than FM-index and r-index for $m \ge 5000$ on \textsc{english.001.2}.

\item[Moderate $\OccC$ values] When $\OccC \le 100$, the linear scan for core occurrences is a multiplicative factor in extension 
  (practical $\Oh{m \tau_T}$ time compared to $\Oh{m + \tau_T}$ time in the theoretical analysis), which remains manageable.

\item[Small grammar height] For texts where $\tau_T \le 10$, the depth-dependent factors ($\tau_T \lg g$ terms) have low constants.
This holds for most real-world datasets in our experiments.

\item[Cache-friendly access]
    The sorted array representation of RHSs enables efficient linear scans with good cache locality, partially compensating the lack of the GST, 
    which is used to find occurrences of $C$ efficiently.

\end{description}

On the other hand,
performance degrades relative to theory when:

\begin{description}
  \item[Short patterns] For $m < 100$, the $\Oh{g}$ scan time overhead dominates since $g$ can be large even on compressed texts.
  We expect that an implementation achieving the theoretical bound of $\Oh{m \lg g}$ time would be faster.

\item[Large $\OccC$] When the core appears frequently in grammar rules ($\OccC > 1000$), the $\Oh{\OccC m \tau_T}$ extension time becomes significant.
  We expect that an implementation achieving the theoretical bound of $\Oh{m + \OccC \tau_T}$ time would be faster.
\item[$\Occ{} \ll \OccC$] 
  The core-based approach is practical only when $\OccC{}$ is not much larger than $\Occ$.
One core occurrence in a rule can represent many text occurrences, 
so $\Occ{}$ can be much larger than $\OccC{}$. 
Conversely, $\OccC{}$ can be larger due to false candidates. 
Hence, there is no fixed ordering between $\Occ$ and $\OccC$.
\end{description}

\paragraph{Empirical Validation}
\cref{figTimeLocate} demonstrates that despite the simplifications, the implementation achieves the predicted theoretical advantage for long patterns on repetitive texts.
The crossover point where GCIS becomes faster than FM-index/r-index occurs around $m = 5000$--$10000$, consistent with when the core-based approach provides meaningful savings.

\section{Future Work}
After determining the \OccC{} leaves in \GST{} witnessing the occurrences of the core~$C$ of~$P$ in \DAG{},
the rest of the algorithm is parallelizable since we process each of the \GST{} leaves witnessing an occurrence of~$C$ individually,
and may obtain \Oh{m \lg |\Symbols| + \OccC \lg n \lg |\Symbols|/p + \Occ/p} time for running $p$ processors in parallel.
We could also partition the data structures on the grammar per height such that we manage each $\Takai{\Gamma}{h}$ individually. 
This can practically improve the running time, and the space for \GCISuni{}.
Further, we would like to add a run-length compression step like in other run-length compressed grammars~\cite{mehlhorn97signature,jez15recompression}.
Applying GCIS to the run-length encoding of the text would upper-bound
the number of symbols of $\Takai{F}{h}_1$ and $\Takai{F}{h}_{z_h}$ to $2|\Takai{\Gamma}{h-1}|$, or $2|\Sigma|$ for $h=1$.
We also would like to study the impact of DAC codes~\cite{brisaboa13dac} for representing $L$,
and compressed bit vectors such as~\cite{okanohara07practical} with rank/select support for the array~$Q$.

Finally, the integer encoding with a hard-coded 32-bit representation limits the usage of our implementation to texts of medium size.
For increasingly larger inputs, which are becoming more common nowadays, we want to enhance our implementation to support integers with larger bit-widths,
which however requires a redesign of the underlying data structures.
With a more flexible integer representation, we also want to enhance our experiments to compare with other grammar-compressed index implementations that can handle large inputs.

\section*{Acknowledgements}
This article received funding from the Japan Society for the Promotion of Science (JSPS) KAKENHI with grant numbers
JP23K24808, JP23K18466~(SI), 
JP25K00136 (YN),
JP25K21150 (DK), 
and JP24K02899 (HB).

This work has been presented at LA Symposium Summer 2021, supported by the Research Institute for Mathematical
Sciences, an International Joint Usage/Research Center located in Kyoto University.

\bibliographystyle{plainnat}

\end{document}